\documentclass[manuscript, screen]{acmart}

\begin{document}

\title{A Survey of Timing Variability in Microservice-Based Software-Defined Vehicles}

\author{Cyrus K. Vattes}
\email{c.k.vattes@tue.nl}
\orcid{0009-0004-1150-9037}
\affiliation{%
    \institution{Eindhoven University of Technology}
    \city{Eindhoven}
    \state{North Brabant}
    \country{Netherlands}
}

\author{Habib Mostafaei}
\affiliation{%
    \institution{Eindhoven University of Technology}
    \city{Eindhoven}
    \state{North Brabant}
    \country{Netherlands}
}
    
\author{Nirvana Meratnia}
\affiliation{%
    \institution{Eindhoven University of Technology}
    \city{Eindhoven}
    \state{North Brabant}
    \country{Netherlands}
}

\renewcommand{\shortauthors}{Vattes et al.}
\begin{abstract}

    Microservice-based systems are modular and adaptable, but their distributed structure makes their timing behavior difficult to analyze and guarantee. Because latency emerges from interactions among service dependencies, shared resources, and coordinating middleware, local timing disturbances can propagate into system-level effects. This survey examines the sources, propagation mechanisms, and observable impacts of timing variability in microservice-based systems, using software-defined vehicles as a motivating example. It compares major classes of timing models by their assumptions about workload stability, execution structure, resource sharing, observability, and guarantee admissibility, and summarizes their limitations within microservice-based software-defined vehicles.

\end{abstract}




%% Rights management information.  This information is sent to you
%% when you complete the rights form.  These commands have SAMPLE
%% values in them; it is your responsibility as an author to replace
%% the commands and values with those provided to you when you
%% complete the rights form.
\setcopyright{acmlicensed}
\copyrightyear{2026}
\acmYear{2018}
\acmDOI{XXXXXXX.XXXXXXX}
%% These commands are for a PROCEEDINGS abstract or paper.
\acmConference[Conference acronym 'XX]{Make sure to enter the correct
  conference title from your rights confirmation email}{June 03--05,
  2018}{Woodstock, NY}
%%
%%  Uncomment \acmBooktitle if the title of the proceedings is different
%%  from ``Proceedings of ...''!
%%
%%\acmBooktitle{Woodstock '18: ACM Symposium on Neural Gaze Detection,
%%  June 03--05, 2018, Woodstock, NY}
\acmISBN{978-1-4503-XXXX-X/2018/06}

%% Use this when submitting an article to a sponsored event. You'll
%% receive a unique submission ID from the organizers
%% of the event, and this ID should be used as the parameter to this command.
%%\acmSubmissionID{123-A56-BU3}

\maketitle

\section{Introduction}

    Software systems increasingly adopt microservice architectures for their modularity, maintainability, and adaptability at runtime \cite{wan2018ApplicationDeploymentUsing, wang2024AutothrottlePracticalBiLevel}. By decomposing applications into loosely-coupled services, components may be deployed, updated, replicated, or replaced independently. However, this decomposition fundamentally changes how execution timing evolves across the system. Work is no longer performed within a single monolithic execution context, but is distributed across services, virtualization environments, networks, and coordination middleware. As a result, the timing behavior of microservice-based systems depends on how those components interact across their execution paths.

    The shift toward microservice-based software systems complicates the assumptions underlying many traditional timing and schedulability models, which generally assume stable execution environments and predictable traffic patterns \cite{leboudec2001NetworkCalculus, stankovic1995ImplicationsClassicalScheduling, kannan2019GrandSLAmGuaranteeingSLAs}. The execution conditions in microservice deployments are often dynamic, partially observable, and governed by multiple interacting control mechanisms \cite{fan2013ModelingOptimizingResource, fan2016FormalAspectOrientedMethod}. Resource contention, asynchronous communication, adaptive scaling, and distributed coordination all alter timing behavior at runtime, making end-to-end timing difficult to infer from local behaviors alone \cite{camilli2018ZonebasedFormalSpecification, laclau2025EnhancingAutomotiveUser}. Existing surveys address parts of this problem, but the interactions among sources of timing variability and their propagation mechanisms have not yet been holistically studied or synthesized.
    
    Software-defined vehicles (SDVs) provide a useful lens for studying these challenges, because they concentrate many of the structural and temporal characteristics of modern distributed systems into a single platform. Vehicle functionality is increasingly implemented as service-oriented software stacks and deployed across heterogeneous compute nodes, networking infrastructures, and abstraction layers. These systems simultaneously require modularity, adaptability, and strong guarantees for safety-critical functions. Consequently, SDVs highlight the growing tension between dynamic, distributed environments and the strict timing requirements of real-time systems.
    
    This survey examines timing behavior and guarantees in microservice-based systems, with an emphasis on SDVs and related mixed-criticality systems. Rather than treating execution timing as a purely local property that accumulates across isolated services, we frame timing behavior as an emergent property of distributed execution, propagation, contention, and control. The contributions of this survey are twofold: first, we explore how timing variability emerges, propagates, and amplifies in modern microservice architectures; and second, we compare the assumptions and limitations of existing timing models when applied to dynamic, heterogeneous, and partially observable microservice-based systems.
    
    The remainder of this survey is organized as follows: Section \ref{sec:background} introduces the background concepts needed to understand the rest of the paper; Section \ref{sec:sources_of_timing_variability} categorizes sources of timing variability at the local, network, and middleware levels; Section \ref{sec:timing_propagation} discusses how timing effects propagate and amplify across system structures; Section \ref{sec:observable_timing_behavior} discusses the observable behaviors resulting from propagated timing effects; Section \ref{sec:timing_models} reviews the families of existing timing models, their assumptions, and their guarantees; and Section \ref{sec:limitations} discusses the limitations of those timing models.

\section{Background}\label{sec:background}

    This section introduces the core concepts and terminology used throughout the survey. Because the timing behaviors presented in this paper emerge from interactions between system components, an explanation of those components is necessary to understand the information presented later. We begin by defining the architectural and operational concepts pertaining to microservice-based systems. Then, we introduce the timing metrics used to characterize execution in those systems. Finally, we describe the structural mechanisms through which timing effects propagate across a system.

\subsection{Core Concepts}

    \subsubsection{Distributed Systems}

        When discussing distributed systems in this survey, we refer to systems whose behaviors emerge from multiple interacting components communicating across a network, with no single component having a complete view of the whole system. This definition covers many system designs where the behavior of an individual node is separate from global behavior, including distributed real-time systems \cite{camilli2018ZonebasedFormalSpecification}, computing continuums \cite{dustdar2023DistributedComputingContinuum}, and microservice-based systems \cite{zhou2018DeltaDebuggingMicroservice, wan2018ApplicationDeploymentUsing}. Importantly, the global behavior in these systems depends on the ordering and coordination of discrete events, as runtime states may be inconsistent across components \cite{vanderaalst2012ProcessMiningOverview, camilli2018FormalFrameworkSpecifying, zhou2018DeltaDebuggingMicroservice}. As a result, the behavior of distributed systems is emergent, and not simply the sum of individual computations \cite{camilli2018ZonebasedFormalSpecification}.

    \subsubsection{Microservice Architectures}

        This survey focuses specifically on microservice-based systems as a service-oriented subset of distributed systems. Service-oriented architecture (SOA) is a system design style in which functional components are connected via explicit, implementation-independent interfaces, through which they consume and provide service capabilities \cite{teixeira2025DeterministicReliableSoftwareDefined, fraccaroli2023TimingPredictabilityIPbased}. Microservice architectures extend this principle by decomposing applications into small, loosely-coupled services that execute in isolation and communicate through shared interfaces \cite{grewal2025SafeTreeExpressiveTree, liu2022ModellingAnalysingReliability, wen2024VirtualizationMicroserviceArchitecture}. In this paper, we use component as a general term for a functional unit, and microservice for an independently-deployable unit of a service; architecture refers to the design philosophy and operational structure of a system (rather than the underlying hardware topology). Microservices are highly modular, and can typically be deployed, updated, replicated, or replaced independently \cite{wan2018ApplicationDeploymentUsing}.
        
        However, service decomposition also fundamentally changes the behaviors and properties of a system. Because applications are segmented into many interacting microservices, computation is distributed across multiple nodes, rather than confined to a single monolithic context \cite{kannan2019GrandSLAmGuaranteeingSLAs, zhou2018DeltaDebuggingMicroservice, lin2018MicroscopePinpointPerformance}. This results in dynamic systems whose behaviors are influenced by application logic, placement decisions, infrastructure, and communication patterns \cite{zhou2018DeltaDebuggingMicroservice, soldani2023AnomalyDetectionFailure}.

    \subsubsection{Execution and Timing}
    \label{sssec:exec_and_time}
    
        To discuss the timing behavior of microservice-based systems, we must first define execution in broader terms than isolated task completion. In this survey, execution refers to the realized runtime progression of an application-level function through invocation, processing, communication, queuing, coordination, and completion across one or more microservices. In practice, requests often spend substantial time outside active computation, so this definition also includes time a request spends in transit, delayed by resource contention, queued at a service boundary, or awaiting completion of another task \cite{kannan2019GrandSLAmGuaranteeingSLAs, fan2013ModelingOptimizingResource}. This matters in distributed systems, where progress can be decoupled from the original order of requests by queued or asynchronous messages \cite{fan2013ModelingOptimizingResource, zhou2018DeltaDebuggingMicroservice}. With this framing, execution is not limited to continuous processor activity, but denotes the end-to-end realization of service functionality through computation, communication, waiting, and coordination. For example, the obstacle detection workflow shown in Figure \ref{fig:BCK_exec_env} spans sensor hardware, heterogeneous communication, multiple microservices, hardware gateways, and actuation stages distributed across multiple execution contexts. We distinguish end-to-end execution from local execution, which refers to execution occurring within a specific service instance and its associated runtime context (e.g., container, virtual machine, ECU, or compute node).

        Where execution describes the progression of work, timing describes the temporal properties of that progression: when work occurs, how long it takes, and whether deadlines or other operational constraints are met. This is especially important in concurrent, embedded, and real-time systems, where even small variations in timing can have significant behavioral consequences \cite{lee2008CyberPhysicalSystems}. In such systems, correctness depends not only on the results of execution, but on when those results are produced \cite{lee2008CyberPhysicalSystems, camilli2018ZonebasedFormalSpecification, fan2016FormalAspectOrientedMethod}. Because common processor architectures, programming languages, and network infrastructures generally do not provide their own timing guarantees, these properties must be modeled and enforced explicitly at the system level rather than assumed from the underlying components \cite{lee2008CyberPhysicalSystems}.
        
    \subsubsection{Observability and Telemetry}\label{sssec:obs_and_telem}

        In distributed microservice architectures, the full behavior of the system cannot be directly observed from any single vantage point. Although architectural models describe how services are expected to interact, the actual runtime behavior of a system emerges dynamically across multiple infrastructure and communication layers. Observability is therefore an epistemic problem, where internal system states must be inferred from externally visible behaviors \cite{otero2024LightweightDistributedTelemetry, ashok2024TraceWeaverDistributedRequest}. The data required to infer local states are inherently fragmented across multiple components of varying granularity, and specific execution details are typically opaque outside individual components \cite{ashok2024TraceWeaverDistributedRequest, gu2023TrinityRCLMultiGranularCodeLevel}. As a result, accurate observation relies on embedding instrumentation into application components to transmit execution and context information across service boundaries \cite{ashok2024TraceWeaverDistributedRequest, otero2024LightweightDistributedTelemetry}. For example, only application-level components can directly track which incoming requests spawn which outgoing requests; making this information available to communication frameworks requires explicit modifications to the application code \cite{ashok2024TraceWeaverDistributedRequest}.

\begin{figure}
            \centering
            \includegraphics[width=0.5\linewidth]{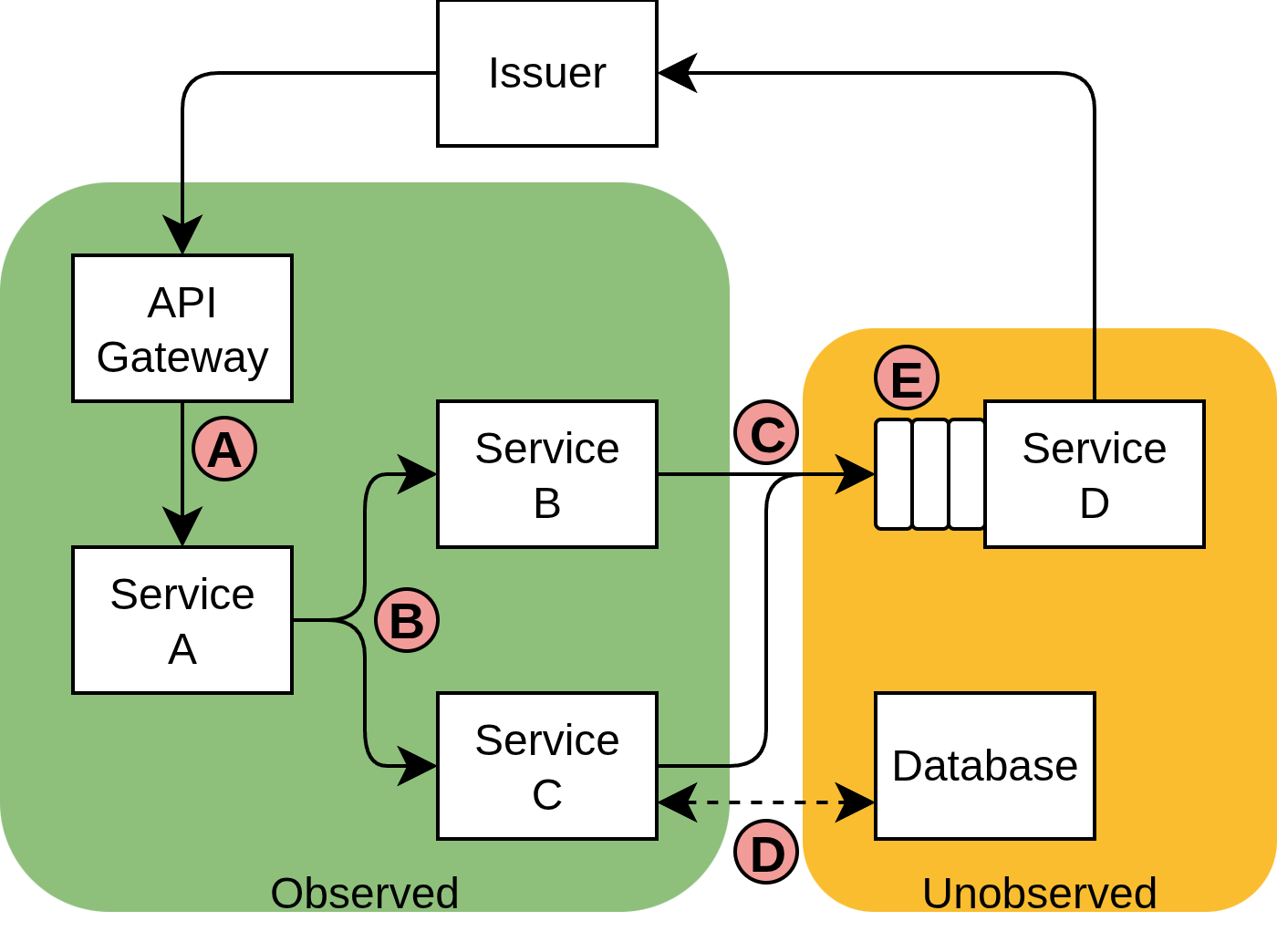}
            \Description{A partially-observable execution path consisting of sequential, fan-out, fan-in, asynchronous, and queued segments.}
            \caption{A partially-observable execution path consisting of \textcircled{A} sequential, \textcircled{B} fan-out, \textcircled{C} fan-in, \textcircled{D} asynchronous, and \textcircled{E} queued segments.}
            \label{fig:BCK_obs_telem}
        \end{figure}

        The runtime evidence for inferring distributed system behavior is provided by telemetry, which is typically organized into traces, metrics, and logs \cite{almaruf2022UsingMicroserviceTelemetry, otero2024LightweightDistributedTelemetry}. Logs record timestamped discrete events, metrics aggregate numerical summaries of system state over time, and traces follow operations associated with individual requests as they move through the system \cite{almaruf2022UsingMicroserviceTelemetry, gu2023TrinityRCLMultiGranularCodeLevel, cornacchia2026ObservabilityEatingYour}. Distributed tracing is the primary method of tracking execution across components; correlated operational spans are assembled into cross-component dependency graphs \cite{ashok2024TraceWeaverDistributedRequest, soldani2023AnomalyDetectionFailure, yu2021MicroRankEndtoEndLatency}. The telemetry spans in Figure \ref{fig:BCK_obs_telem} are only partially observable: they may include incomplete spans or dropped spans, unobserved interactions, and asynchronous calls. These partially-observed spans can still be assembled into approximate execution paths, but the resulting graph will be incomplete. These reconstructed execution graphs are essential for reconciling the gap between the declared structure and the actual runtime behavior of the system \cite{almaruf2022UsingMicroserviceTelemetry}. However, because they are assembled from partial observations and inference, they are not perfect mirrors of the system at runtime. In a very real sense, reconstructed execution graphs are hybrid objects whose accuracy depends on instrumentation, sampling strategy, and context propagation \cite{ashok2024TraceWeaverDistributedRequest, cornacchia2026ObservabilityEatingYour, gu2023TrinityRCLMultiGranularCodeLevel}.
        
    \subsubsection{Software-Defined Vehicles}

        Software-defined vehicles (SDVs) concentrate many of the structural and temporal challenges of distributed systems into a single platform. This makes them a good use-case for examining the specific properties of microservice-based systems, and we will refer back to them throughout this paper.
        
        Vehicle manufacturers are moving away from traditional Electronic/Electrical (E/E) architectures toward centralized and zonal architectures built around high-performance computers (HPCs), zonal controllers, and Ethernet-based communication backbones \cite{fraccaroli2023TimingPredictabilityIPbased, teixeira2025DeterministicReliableSoftwareDefined, mauser2026MixedCriticalitySoftwareArchitectures, pan2024SoftwareDefinedVehiclesModelBased}. This is accompanied by a broader shift toward software-driven designs that decouple hardware from software functionality, improve modularity, and allow for over-the-air updates \cite{wen2024VirtualizationMicroserviceArchitecture, laclau2025EnhancingAutomotiveUser}. In this setting, vehicle functions are increasingly deployed as software services that interact over middleware and shared infrastructure, rather than as fixed workflows between Electronic Control Units (ECUs).

\begin{figure}
            \centering
            \includegraphics[width=0.7\linewidth]{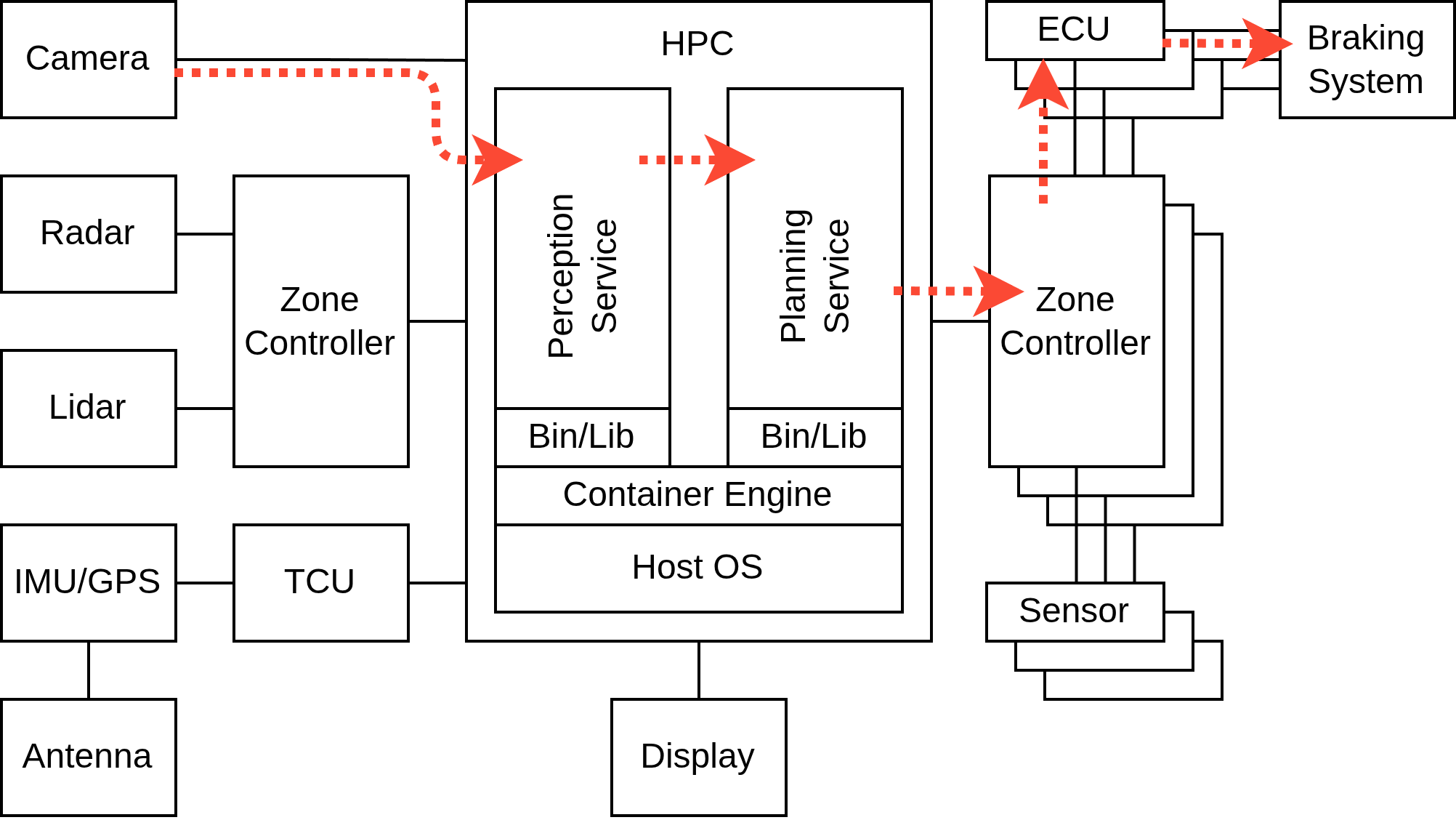}
            \Description{An example workflow for obstacle detection and automatic braking in a software-defined vehicle.}
            \caption{An example workflow for obstacle detection and automatic braking in a software-defined vehicle.}
            \label{fig:BCK_exec_env}
        \end{figure}

        The SDV execution environment is organized into abstraction layers that simplify communication between services and the underlying infrastructure. At the lowest layers, hardware virtualization allows multiple operating systems to coexist on a single ECU, while hardware abstraction layers decouple services from hardware-specific implementation \cite{laclau2024DesignDynamicArchitectures, teixeira2025DeterministicReliableSoftwareDefined}. Within the HPC or ECUs, software virtualization allows multiple applications or execution environments to share the same hardware. Above this, network virtualization enables transport-agnostic communication and the use of deterministic networking protocols \cite{laclau2024DesignDynamicArchitectures, teixeira2025DeterministicReliableSoftwareDefined}. At the top, observability and orchestration mechanisms coordinate scheduling, resource allocation, and lifecycle management decisions across the system. Figure \ref{fig:BCK_exec_env} shows a simplified diagram of an SDV; hardware components are connected by communication buses, and software components are virtualized within the HPC.
        
        This layered approach improves flexibility and enforces spatial and temporal isolation of services, allowing heterogeneous workloads to safely coexist \cite{ferraro2023TimesensitiveAutonomousArchitectures, holstein2015ContradictionSeparationVirtualization, mauser2026MixedCriticalitySoftwareArchitectures}, but it also increases operational complexity. When applications span multiple abstraction layers, cross-layer interactions and environmental uncertainties make system behavior difficult to predict \cite{camilli2018ZonebasedFormalSpecification, laclau2025EnhancingAutomotiveUser}. Operating system concurrency, virtualization overhead, and resource contention all introduce nondeterminism that cannot be fully eliminated by scheduling policies \cite{lee2008CyberPhysicalSystems, zhao2020RhythmComponentdistinguishableWorkload}. Because SDVs combine service decomposition, heterogeneous execution environments, and strict safety requirements, this survey will use them as a lens for exploring the timing behaviors of microservice architectures. The SDV execution environment in Figure \ref{fig:BCK_exec_env} includes a central HPC, distributed zonal controllers, ECU gateways, heterogeneous infrastructure, and mixed-criticality workflows. One such workflow---obstacle detection and collision avoidance---is highlighted, and will serve as a running example throughout this survey.

\subsection{Timing Metrics}

    \subsubsection{Latency}

        We refer to the elapsed time interval between the arrival or generation of a request and the availability of the corresponding result as latency \cite{leboudec1998ApplicationNetworkCalculus}. This reflects both the responsiveness of individual services and the delays introduced by communication and coordination mechanisms through which those services interact \cite{kannan2019GrandSLAmGuaranteeingSLAs}. When necessary, we distinguish between execution latency (the time elapsed during execution within a specific runtime context, see Section \ref{sssec:exec_and_time}) and delay (the time spent outside execution, such as during network transmission or while queued).

        Because latency in a distributed system may not be entirely local or sequential, we must also define an execution path through such a system. An execution path is the ordered runtime structure through which a request is realized, from arrival to end result \cite{grewal2025SafeTreeExpressiveTree, kannan2019GrandSLAmGuaranteeingSLAs}. Such a path may include local execution within one or more service instances, communication between services, queuing at service boundaries, and so on. Although the term suggests a linear sequence, execution paths in microservice architectures may include branching, parallel, or asynchronous stages. The relevant execution path consists of the ordered causal work that contributes to the completion of a request, and must often be reconstructed from observed behavior \cite{ashok2024TraceWeaverDistributedRequest}. Execution path reconstruction is discussed in a later section.
        
        \begin{figure}
            \centering
            \includegraphics[width=0.7\linewidth]{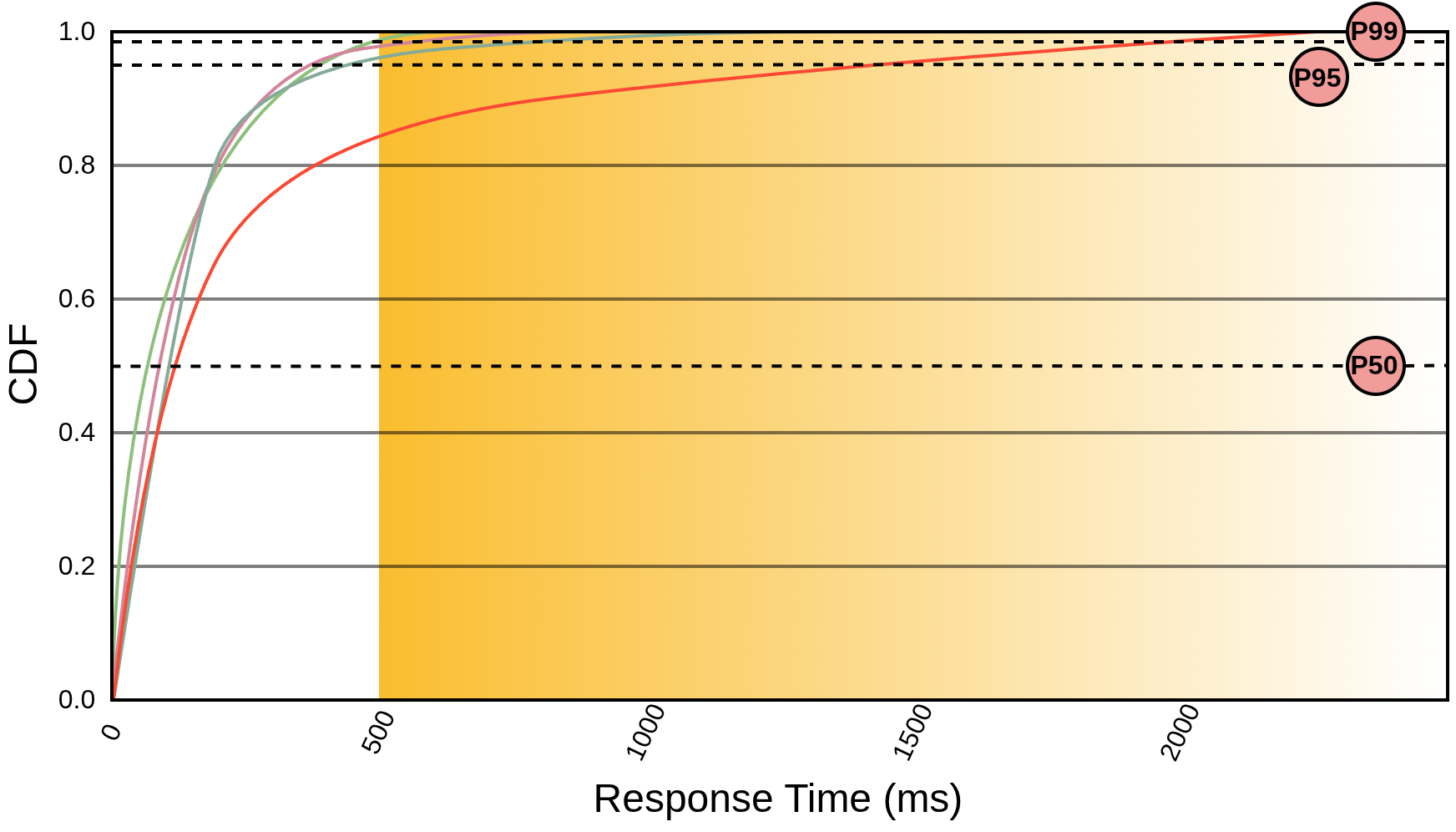}
            \Description{Latency distributions for multiple distributed applications; the 99th percentile latency has a particularly long tail.}
            \caption{Latency distributions for multiple distributed applications \cite{luo2022InDepthStudyMicroservice}.}
            \label{fig:BCK_latency_dist}
        \end{figure}

        For the purposes of this survey, we distinguish between latency at two levels of abstraction. Local execution latency refers to the execution time for a single service instance within its runtime context. End-to-end (or global) latency refers to the total elapsed time accumulated across the entire execution path. In microservice architectures, this path commonly spans multiple services, network segments, middleware layers, and coordination points \cite{kannan2019GrandSLAmGuaranteeingSLAs, sperling2024ReducingCommunicationCost}. Because the overhead incurred at service boundaries across the execution path is difficult to predict, inferring end-to-end latency from local execution latency is nontrivial \cite{fraccaroli2023TimingPredictabilityIPbased, choi2024EnhancingRecoveryPerformance, park2023AnalysisIPCANCommunication}. Indeed, latency in distributed systems is often unevenly distributed, and rare high-latency events disproportionately affect observed system behavior. In Figure \ref{fig:BCK_latency_dist}, all applications will experience SLO violations in at least 1\% of requests, even with a relatively generous timing allowance. End-to-end latency is therefore a structurally-dependent measure of timing behavior, influenced by how efficiently information can propagate through the system \cite{teixeira2025DeterministicReliableSoftwareDefined, sperling2024ReducingCommunicationCost}.

    \subsubsection{Timing Variability}

        The deviation of observed timing from an expected delay or timing pattern is referred to as jitter. In networking and control settings, jitter is commonly used to describe variance in packet delivery, as transmissions are expected to follow regular arrival curves \cite{leboudec1998ApplicationNetworkCalculus, leboudec2001NetworkCalculus}. These variations typically stem from fluctuations in the network environment, such as those caused by scheduling policies, queuing effects, protocol translation, or imperfect clock synchronization \cite{leboudec2001NetworkCalculus, fraccaroli2023TimingPredictabilityIPbased, teixeira2025DeterministicReliableSoftwareDefined}. Jitter is most important in settings when consistency matters as much as latency, such as when periodic feedback and control loops are used. In these cases, irregular timing can interfere with service coordination, even when the average delay remains acceptably bounded \cite{fraccaroli2023TimingPredictabilityIPbased, ferraro2023TimesensitiveAutonomousArchitectures}. The long latency tail in Figure \ref{fig:BCK_latency_dist} is one such manifestation of timing irregularities with a bounded average delay.

        Although jitter is a useful concept, its precise networking definition is too narrow for the scope of this survey. Instead, we use timing variability to refer to deviations in timing behavior across the full execution path, including but not limited to classical jitter. This broader term is needed because timing fluctuations in distributed microservice architectures can have many causes unrelated to communication or periodicity. For example, resource contention can lead to queuing effects that vary with traffic volume, network conditions, or execution environments \cite{kannan2019GrandSLAmGuaranteeingSLAs, lee2008CyberPhysicalSystems, zhao2020RhythmComponentdistinguishableWorkload, soldani2023AnomalyDetectionFailure}. These effects are often nondeterministic and heterogeneous across service components and execution paths \cite{zhao2020RhythmComponentdistinguishableWorkload, kamboj2025LeveragingPetriNets}. It is imperative for timing-critical sensing and control guarantees to tolerate timing fluctuations rather than assume network or computational regularity \cite{fraccaroli2023TimingPredictabilityIPbased, teixeira2025DeterministicReliableSoftwareDefined}. This survey, therefore, uses timing variability in all cases.

    \subsubsection{Operational Constraints and Violations}

        Although throughput does not directly measure timing behavior, it is essential for understanding how system behavior changes under load. Broadly, it describes the rate at which work is processed relative to incoming demand, and is therefore closely related to network and service capacity \cite{vanderaalst2012ProcessMiningOverview}. In microservice-based systems, throughput depends on traffic rates, resource availability, scaling decisions, and how work is distributed across compute nodes \cite{wan2018ApplicationDeploymentUsing, kannan2019GrandSLAmGuaranteeingSLAs}. When demand approaches or exceeds the effective system capacity, queuing effects can accumulate, impacting timing behavior. This means that, in practice, throughput estimates are useful for detecting bottlenecks and optimizing resource efficiency \cite{vanderaalst2012ProcessMiningOverview, almaruf2022UsingMicroserviceTelemetry}.
        
        Miss rates measure the recurrence of timing violations, or the frequency of jobs, tasks, or requests for which a timing requirement is not satisfied over a period of time. This can be useful for determining whether such violations are isolated events or a recurring pattern that impacts scheduling or resource management \cite{fan2013ModelingOptimizingResource}. Although miss rates are often associated with task-level deadline misses, they are equally relevant to distributed workloads, where scheduling and resource management decisions depend on the accuracy of completion time estimates \cite{zhao2020RhythmComponentdistinguishableWorkload, fan2013ModelingOptimizingResource, fan2016FormalAspectOrientedMethod}.

        At the system level, individual delays or deadline misses often compound into a failure to satisfy service-level commitments. We refer to these system-level failures as contract violations \cite{dustdar2023DistributedComputingContinuum, teixeira2025DeterministicReliableSoftwareDefined, ferraro2023TimesensitiveAutonomousArchitectures}. This survey refers to three such performance agreements: a Service-Level Agreement (SLA) is a commitment between the service provider and the user that addresses specific aspects of the service, such as responsibilities, expected performance level, and reporting requirements; a Service-Level Objective (SLO) is a measurable performance goal of the SLA, such as response time; and Quality of Service (QoS) is the measurable end-to-end performance of a service, which may be guaranteed in advance by an SLA. These commitments define the acceptable behavior for a service or platform \cite{dustdar2023DistributedComputingContinuum, teixeira2025DeterministicReliableSoftwareDefined, ferraro2023TimesensitiveAutonomousArchitectures}.

        In microservice-based systems, these are frequently end-to-end commitments, where satisfying the system-level commitments depends on meeting disaggregated timing requirements across multiple execution stages \cite{kannan2019GrandSLAmGuaranteeingSLAs}. For this reason, contracts for safety-critical or mixed-criticality services are often enforced through resource reservation and distributed QoS policies that constrain both scheduling and communication behavior \cite{laclau2025EnhancingAutomotiveUser, teixeira2025DeterministicReliableSoftwareDefined}.

\subsection{Propagation Structures}

    \subsubsection{Execution Graphs}

\begin{figure}
            \centering
            \includegraphics[width=\linewidth]{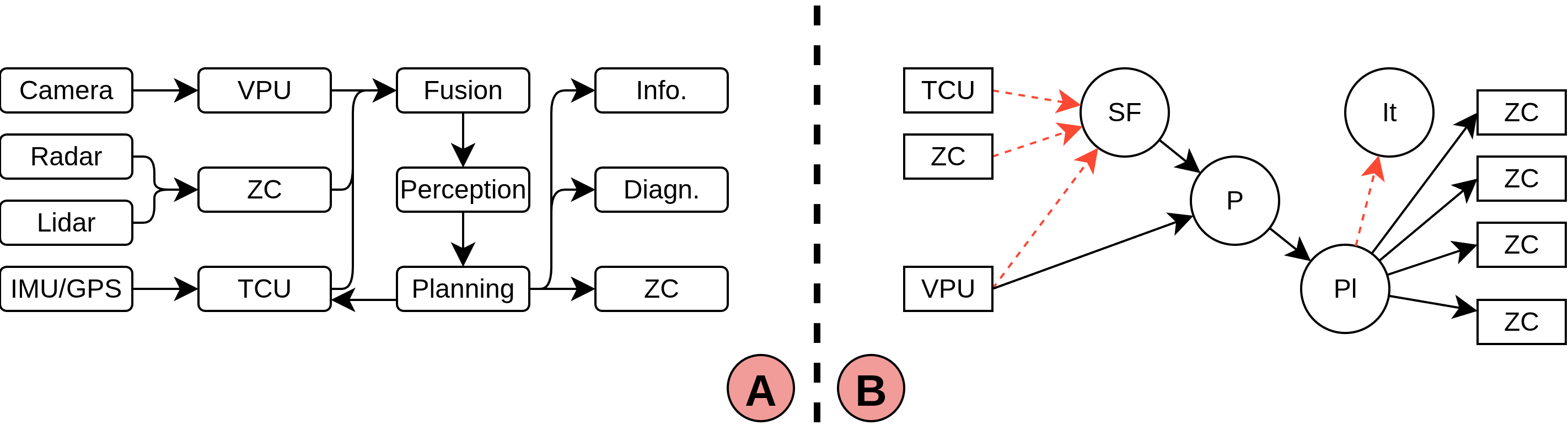}
            \Description{Potential declared and reconstructed execution graphs for an SDV collision avoidance workflow. The two graphs are not identical, and several directed connections are added or removed from the reconstructed graph.}
            \caption{Potential \textcircled{A} declared and \textcircled{B} reconstructed execution graphs for an SDV collision avoidance workflow. When the camera detects an obstacle, the decoder triggers a priority workflow that bypasses multiple services. Links to bypassed services (red) may be lost if reconstruction occurs during the priority workflow.}
            \label{fig:BCK_exec_graphs}
        \end{figure}

        Execution and dependency graphs explicitly model the structural relations through which work is organized in microservice-based systems. They typically represent service operations or execution stages as nodes, with directed edges indicating call, order, or dependency relations \cite{grewal2025SafeTreeExpressiveTree, yu2021MicroRankEndtoEndLatency, ashok2024TraceWeaverDistributedRequest}. There are several common representations for execution graphs, including scheduling graphs, service trees, process trees, Petri nets, and workflow nets, which model different abstraction levels and dependency structures \cite{kannan2019GrandSLAmGuaranteeingSLAs, kamboj2025LeveragingPetriNets, leemans2013DiscoveringBlockStructuredProcess}. We will distinguish between two types of execution graphs: declared (constructed as design-time process or orchestration models) and reconstructed (built from runtime telemetry to approximate dependency relations) \cite{vanderaalst2012ProcessMiningOverview, kamboj2025LeveragingPetriNets, almaruf2022UsingMicroserviceTelemetry}. The reconstructed and declared graphs can diverge significantly if the system is partially observable (see Section \ref{sssec:obs_and_telem}). For example, the runtime graph in Figure \ref{fig:BCK_exec_graphs} may include execution paths not present in the declared graph (VPU $\to$ Perception), while excluding paths not observed by telemetry (e.g., Sensor Fusion $\to$ Perception). The importance of both graph types lies in how accurately they capture the runtime structure of a system, which impacts how timing effects propagate.
        
        Within execution graphs, the composition of execution paths largely determines their relevance to system-level timing behavior. Delay accumulation in parallel or asynchronous paths will depend on how isolated, interactive, or synchronized their stages are \cite{fan2016FormalAspectOrientedMethod, kannan2019GrandSLAmGuaranteeingSLAs}. Shared services (i.e., points where execution graphs converge) are particularly important, because they increase load and create propagation paths between otherwise distinct pathways \cite{kannan2019GrandSLAmGuaranteeingSLAs}. For example, the perception and planning services in Figure \ref{fig:BCK_exec_graphs} may create queuing effects in upstream services if they receive a large volume of high-priority traffic. As a result, local delays can easily spread through implicit propagation paths created by resource contention or co-location (both are common in microservice architectures), even when no direct invocation edge exists \cite{gu2023TrinityRCLMultiGranularCodeLevel, lin2018MicroscopePinpointPerformance, wu2020MicroRCARootCause}. Because of this tendency for delays and interference to spread, later sections treat execution structure as the main mechanism for timing effect propagation.

    \subsubsection{Execution Environments}

        In addition to the logical dependency structure of the service graph, timing behavior is also determined by the environment in which those services run. In microservice architectures, individual services often execute inside containers or virtual machines, potentially nested within layered stacks that span hardware, operating systems, middleware, network abstractions, and coordination layers \cite{grewal2025SafeTreeExpressiveTree, pan2024SoftwareDefinedVehiclesModelBased, laclau2024DesignDynamicArchitectures, teixeira2025DeterministicReliableSoftwareDefined}. These execution environments determine how resources are allocated and shared, how strongly services are isolated from one another, and how requests are transmitted between services. Hypervisor-based virtualization isolates workloads at the hardware abstraction layer, while container-based virtualization isolates them at the operating system level while sharing a common kernel \cite{wan2018ApplicationDeploymentUsing, wen2024VirtualizationMicroserviceArchitecture}. Consequently, the choice of environment influences a wide range of timing behaviors, including service startup time, scheduling and communication overhead, and resource availability \cite{wen2024VirtualizationMicroserviceArchitecture, mauser2026MixedCriticalitySoftwareArchitectures, otero2025ExtensibleLightweightFramework, park2023AnalysisIPCANCommunication}.

        Execution environments also impose further structural conditions that affect timing propagation. Network and deployment topology determine which services are co-located, how services communicate across the network, and which protocol boundaries separate different hosts or gateways \cite{park2023AnalysisIPCANCommunication, laclau2024DesignDynamicArchitectures}. These placement decisions can create or mitigate networking bottlenecks and queuing effects caused by protocol translation \cite{teixeira2025DeterministicReliableSoftwareDefined}. As an example, the placement of the sensor fusion and perception services in Figure \ref{fig:BCK_exec_graphs} contributes to potential queuing behavior and resource contention. Shared infrastructure can create implicit coupling (e.g., contention for compute, memory, or network resources) between services, allowing timing effects to propagate even without direct service invocation \cite{zhang2025NetworkAwareReliabilityModeling, gu2023TrinityRCLMultiGranularCodeLevel, kannan2019GrandSLAmGuaranteeingSLAs}. These interactions are exacerbated in mixed-criticality systems, where workloads with different safety and timing requirements coexist on shared infrastructure, requiring partitioning, network segmentation, bandwidth reservation, fixed-cycle scheduling, traffic shaping, or other isolation mechanisms \cite{mauser2026MixedCriticalitySoftwareArchitectures, ferraro2023TimesensitiveAutonomousArchitectures}. Such measures can reduce harmful coupling between components, but they cannot fully constrain timing propagation in distributed systems \cite{holstein2015ContradictionSeparationVirtualization, whaiduzzaman2021ResilientFogIoTFramework}.

        As a final note, many execution frameworks inhabit a somewhat grey area between execution environment and middleware. For example, ROS (Robot Operating System) provides interfaces and communication protocols that are both structure- and coordination-relevant. For the purposes of this paper, we consider these to be execution environments when their primary focus is on providing runtimes, virtualization, resource provisioning, or (local) scheduling. Cases relating to coordination, control, or observability are discussed in the following section.
        
    \subsubsection{Coordination and Control Middleware}

        Many of the scheduling and routing decisions that influence execution timing in microservice-based systems are managed at the system level by middleware that we collectively refer to as control layers. These include monitoring and coordination mechanisms (e.g., orchestrators, service meshes, or observability frameworks) that monitor a system and govern its operation and lifecycle decisions \cite{wan2018ApplicationDeploymentUsing, pan2024SoftwareDefinedVehiclesModelBased, laclau2025EnhancingAutomotiveUser}. These are typically implemented as middleware and deployed alongside application services. In layered platforms, they may also coordinate hardware, network, and application resources through shared abstractions that help align scheduling and optimization decisions in response to changing system conditions \cite{laclau2024DesignDynamicArchitectures}. Though their form and implementation depends on the functional needs of the system, middleware is always part of the executing structure and timing behavior of the system \cite{laclau2024DesignDynamicArchitectures, wan2018ApplicationDeploymentUsing}.
        
\begin{figure}[hb]
            \centering
            \includegraphics[width=0.39\linewidth]{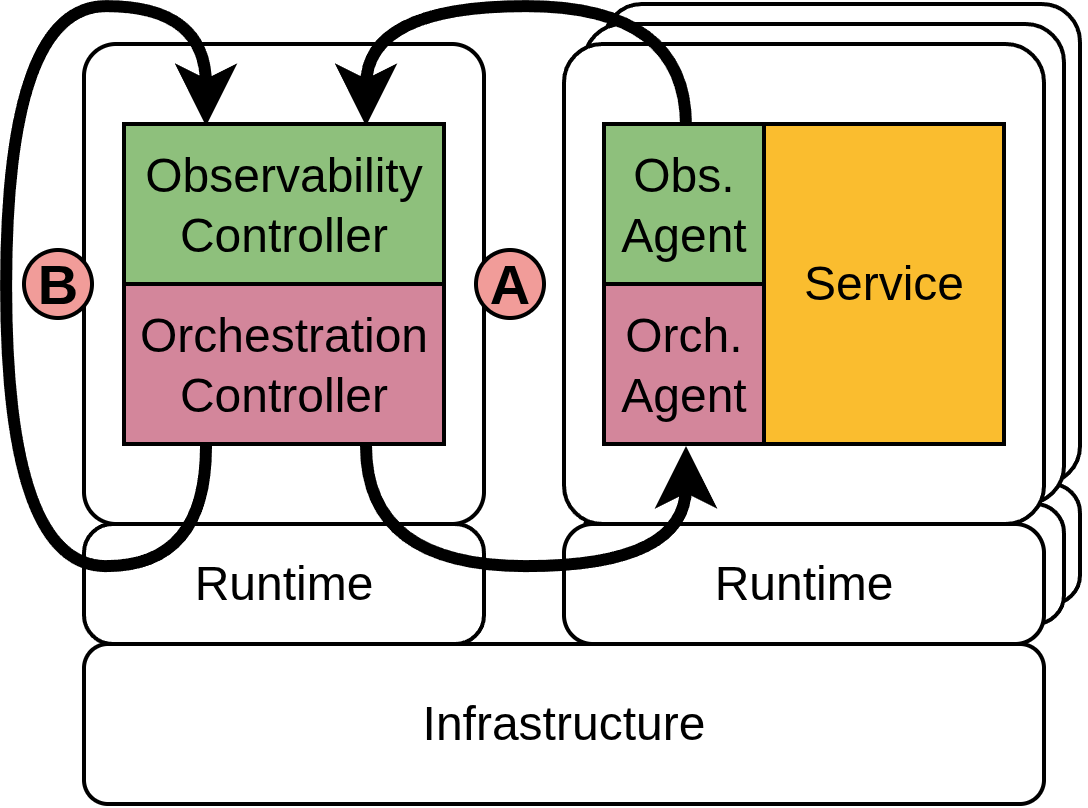}
            \Description{The observability and orchestration control layers form feedback loops within two containers. The orchestration and observability controllers and agents communicated directly with each other, so their feedback loops are largely independent of the application they are running alongside.}
            \caption{Observability and orchestration control layers form feedback loops within application layers (\textcircled{A}) or across abstraction layers (\textcircled{B}) in microservice-based systems.}
            \label{fig:BCK_control_layers}
        \end{figure}

        Because orchestration and observability mechanisms form a feedback loop, they introduce their own timing effects into the system. Figure \ref{fig:BCK_control_layers} highlights the feedback loops for one possible implementation for orchestration and observability layers. Monitoring, coordination, and reconfiguration tasks form dependency cycles where a delay in one process can easily spread to other, seemingly unrelated, processes. The observed system state is often delayed or inaccurate, which incurs delays when control tasks are performed. For example, bottleneck detection becomes both less efficient and less accurate when telemetry is incomplete or coarsely sampled \cite{soldani2023AnomalyDetectionFailure}. This is particularly challenging in heterogeneous networks, where latencies may be nondeterministic \cite{camilli2018ZonebasedFormalSpecification, laclau2025EnhancingAutomotiveUser}, or when a centralized orchestrator assumes that global state will remain continuous and coherent \cite{cornacchia2026ObservabilityEatingYour}.
        
        The monitoring stacks that inform control decisions compete with application workloads for compute, memory, and bandwidth, and can significantly alter response times and tail behavior \cite{cornacchia2026ObservabilityEatingYour, ashok2024TraceWeaverDistributedRequest, wang2020WorkflowAwareAutomaticFault}. Because control layers have nontrivial resource requirements, their performance is profiled under common operating conditions. However, their impacts on timing are often obscured by specific implementation or instrumentation details \cite{cornacchia2026ObservabilityEatingYour, wang2020WorkflowAwareAutomaticFault, soldani2023AnomalyDetectionFailure}.
        
\section{Sources of Timing Variability}\label{sec:sources_of_timing_variability}

    Global timing variability in microservice-based systems largely emerges from interactions between components and abstraction layers. Separating the sources of variability according to their primary execution or interaction contexts gives us a useful way of analyzing their effects and relations. In this survey, we group the sources of timing variability into three categories:
    \begin{enumerate}
        \item those that occur within a single (local) execution context,
        \item those introduced by communication between services,
        \item and those introduced by coordination and control middleware.
    \end{enumerate}
    These categories are not wholly independent of each other, but they provide a structured view of how timing variability enters and propagates across the layers of a distributed system.

\subsection{Local Variability}

    \subsubsection{Scheduling}\label{sssec:scheduling}

        When independent services are composed into larger workflows, scheduling mechanisms determine how compute resources are allocated during execution \cite{fan2013ModelingOptimizingResource}. In microservice-based systems, this scheduling typically occurs within containerized or virtualized environments rather than on bare metal. Although virtualization improves deployment flexibility, it also introduces measurable overhead, including increased application start-up times and reduced scheduling and execution efficiency relative to native deployments \cite{wen2024VirtualizationMicroserviceArchitecture}. As a result, timing behavior at the node level depends on the efficiency of the local execution environment as much as on service logic.

        When multiple services are co-located within the same node or virtual environment, they compete for processor time and memory. This makes local scheduling a direct source of timing variability, as local execution progress depends on how the runtime environment and orchestration layers distribute access to resources. Interference between co-located services remains a persistent source of delay and unpredictability, particularly in high-density deployments \cite{sampaio2019ImprovingMicroservicebasedApplications, wang2024AutothrottlePracticalBiLevel}. Even under a single scheduler or orchestrator, the timing of node-level execution is partially determined by the extent to which workloads interfere with one another.
        
    \subsubsection{Queuing}\label{sssec:queuing}

\begin{figure}
            \centering
            \includegraphics[width=0.6\linewidth]{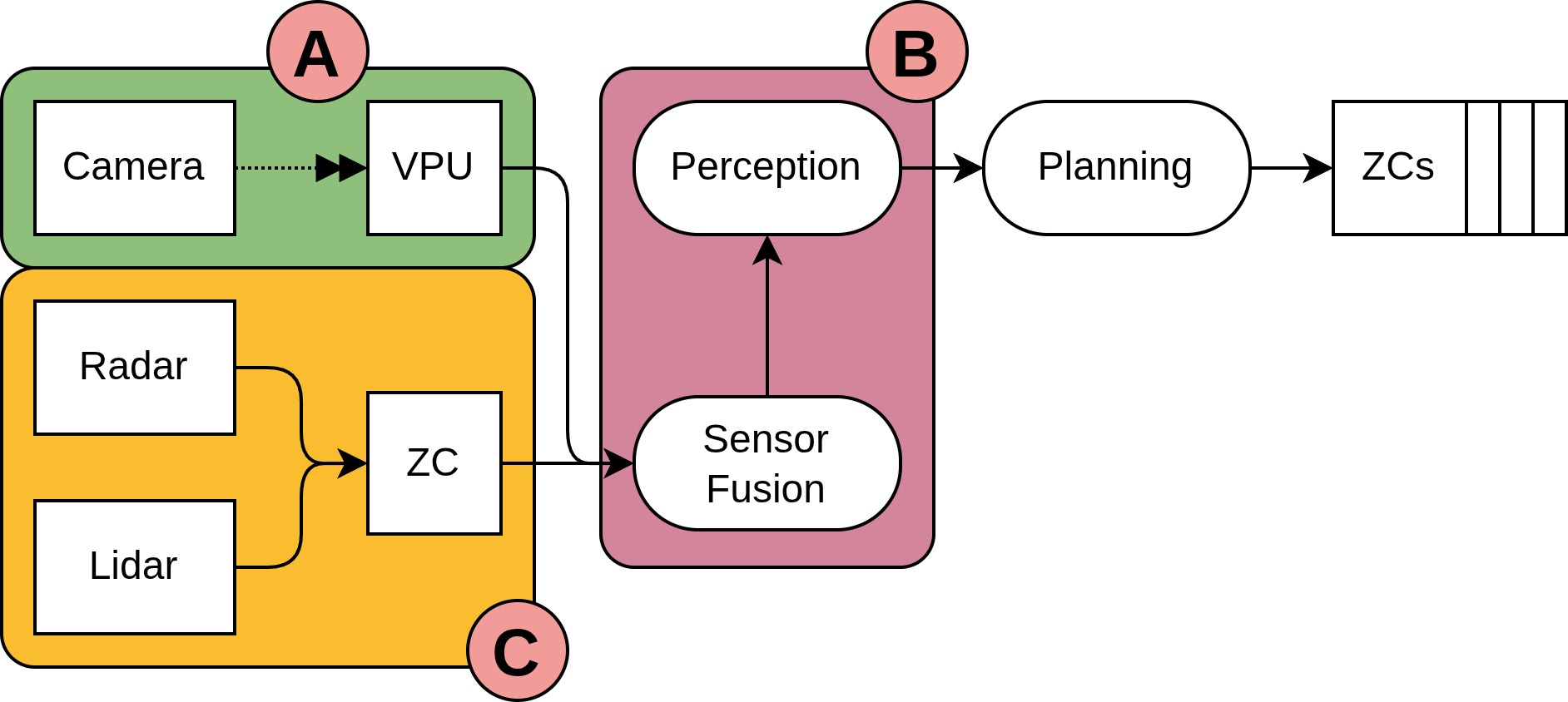}
            \Description{A subsection of an SDV microservice workflow, with potential sources of local timing variability highlighted.}
            \caption{Potential sources of local timing variability in an SDV microservice workflow, including \textcircled{A} queuing caused by high throughput, \textcircled{B} resource contention or scheduling conflicts between co-located services, and \textcircled{C} synchronization errors caused by request fan-in.}
            \label{fig:SRC_local}
        \end{figure}

        Queuing effects provide one of the clearest mechanisms by which local node behavior becomes observable as timing variability. When incoming requests arrive faster than they can be processed, they may be forced to wait in queues until they can be scheduled. Figure \ref{fig:SRC_local} (\textcircled{A}) shows one such scenario, where data streaming from cameras or sensors can form queues if not processed quickly enough. Even when traffic patterns are otherwise normal, the presence of high-priority requests can force lower-priority requests to queue \cite{fan2013ModelingOptimizingResource}. The response time of an individual service thus includes not only its execution time, but also the time spent waiting for access to the processor or other resources.
        
        The number of queued requests increases with traffic volume, and waiting times become both larger and more variable. This causes response time to become increasingly difficult to predict, even when the underlying service is behaving normally \cite{fan2013ModelingOptimizingResource}. Because of this amplifying effect on local execution times, transient traffic patterns can increase the response time of downstream services. In this way, queuing transforms local scheduling or resource availability pressures into directly observable timing variations.
        
    \subsubsection{Resource Contention}

        Resource contention arises when multiple co-located services compete for shared resources. This is generally measured within the node (e.g., for CPU time or memory), though the differences between node-level and network contention (congestion) are largely semantic \cite{sampaio2019ImprovingMicroservicebasedApplications}. Contention is common in microservice architectures because services are usually deployed in containerized or virtualized environments that prioritize flexibility and density over exclusive access to hardware. While such environments support scalable deployment, they also create conditions in which interference between workloads degrades performance \cite{wen2024VirtualizationMicroserviceArchitecture, sampaio2019ImprovingMicroservicebasedApplications}. For example, the performance of the co-located microservices shown in Figure \ref{fig:SRC_local} (\textcircled{B}) might degrade if their resource usage interferes with normal operation.

        This interference persists even when orchestration and resource allocation mechanisms are used to regulate placement and execution. Such mechanisms can reduce harmful interactions, but they do not fully eliminate the timing impact of shared-resource competition \cite{wang2024AutothrottlePracticalBiLevel}. Contention is a fundamental source of single-node timing variability that increases execution time, interacts with local scheduling policies, and feeds directly into queue growth and response time. Before communication is even considered, the timing behavior of a node is already dependent on how co-located workloads share limited local resources.

    \subsubsection{Synchronization Errors}\label{sssec:synchronization}

        In addition to the variability caused by scheduling and interference among sequential tasks, a single node can experience timing variability when synchronizing concurrent tasks. A microservice instance may process many requests simultaneously using, e.g., thread pools. Local execution in parallel or concurrent environments typically results in a partially ordered set of activities whose observed order depends on how they are interleaved at runtime. When testing a concurrent application (or microservice), each observation captures only one possible interleaving of start and end times for the parallel tasks. In these cases, variations in processing time (e.g., local system delays or network latency) can make event ordering nondeterministic \cite{kamboj2025LeveragingPetriNets}. This can be problematic if the output of a service depends on the correct ordering of events, as in Figure \ref{fig:SRC_local} (\textcircled{C}). Synchronization, or the lack thereof, then becomes a source of timing variability, even in otherwise synchronous local executions \cite{lee2008CyberPhysicalSystems}.

        Asynchronous invocations reduce blocking behavior, but may introduce relaxed ordering. Callbacks, message queues, and non-blocking execution may produce event matching errors and result in incorrect estimates of request transit time \cite{zhao2020RhythmComponentdistinguishableWorkload, zhou2018DeltaDebuggingMicroservice}. Without proper coordination, these reordered interactions may lead to misaligned timing observations or service failures that can propagate to downstream contexts \cite{wang2023ComplexBehavioralInteraction, zhou2018DeltaDebuggingMicroservice}.
        
        Because the ordering of events often serves as a basis for timing analysis, synchronization is especially important in microservice-based systems. However, synchronization mechanisms (e.g., locks, priority inversion, etc.) can conflict with timing formalisms and introduce behavior that is difficult to bound \cite{lee2008CyberPhysicalSystems}, and complex asynchronous execution paths increase the number of possible interleavings and make synchronization errors more likely \cite{wang2023ComplexBehavioralInteraction}. These concerns are one main reason for the adoption of concurrency-aware execution models, such as Petri nets (see Section \ref{ssec:state-based_models}).

\subsection{Network Variability}

    \subsubsection{Network Delays and Interference}

        The distributed services of microservice and service-oriented architectures typically communicate over network interfaces using patterns such as request-response and publish-subscribe \cite{wen2024VirtualizationMicroserviceArchitecture, fraccaroli2023TimingPredictabilityIPbased}. When deployments span heterogeneous networks, each component contributes different timing characteristics \cite{whaiduzzaman2021ResilientFogIoTFramework, sperling2024ReducingCommunicationCost}. This results in communication becoming a major source of timing variability in networked systems. Unlike single-node variability, which arises primarily from local scheduling and resource sharing decisions, network variability also includes delays introduced by data transmission and translation between services.

        There are multiple sources of communication delay, including transmission time, protocol and processing overhead, routing delays, and per-hop forwarding delays \cite{do2024PerformanceAnalysisTraffic, park2023AnalysisIPCANCommunication}. When multiple traffic flows share the same network infrastructure, these baseline delays are further amplified by interference, queuing, and contention effects \cite{do2024PerformanceAnalysisTraffic, kannan2019GrandSLAmGuaranteeingSLAs}.

        Several mechanisms exist for constraining network variability. Perhaps the most relevant to SDVs and real-time systems is Time-Sensitive Networking (TSN), which introduces synchronization, traffic shaping, and scheduling mechanisms that allow communication latency to be bounded \cite{teixeira2025DeterministicReliableSoftwareDefined, fraccaroli2023TimingPredictabilityIPbased}. However, network interactions are difficult to control without complex deterministic networking stacks. Even under synchronized and carefully managed conditions, communication remains sensitive to interference and congestion \cite{do2024PerformanceAnalysisTraffic, fraccaroli2023TimingPredictabilityIPbased}.
        
    \subsubsection{Network Failure}

        Unstable network conditions introduce timing overhead when recovery or resilience mechanisms consume system resources \cite{choi2024EnhancingRecoveryPerformance, whaiduzzaman2021ResilientFogIoTFramework}. For example, retry loops consume compute cycles and repeatedly send liveness requests over the network. In practice, this means that network failures not only affect timing variability by their incurred communication costs, but also by whatever mechanisms are in place to detect and recover from disruptions, including tolerance policies. These mechanisms increase end-to-end latency beyond what is expected under normal operating conditions.

        This may become problematic in safety-critical settings, where communication delays are often included as part of operational contracts. If an autonomous workflow must execute within a strict deadline of 100 milliseconds, including sensing, communication, and downstream computation, then any unexpected network instability directly reduces the time budget available for those downstream execution stages \cite{sperling2024ReducingCommunicationCost}. Network failures or instability therefore contribute additional overhead that is difficult to predict or observe unless specifically measured by observability middleware.

    \subsubsection{Congestion}

        Congestion, or when multiple traffic flows compete for limited transmission capacity, is a major source of timing variability in networks. When network infrastructure is shared, the contention creates queuing delays that increase both average latency and latency variance \cite{do2024PerformanceAnalysisTraffic, kannan2019GrandSLAmGuaranteeingSLAs}.
        
\begin{figure}
            \centering
            \includegraphics[width=0.5\linewidth]{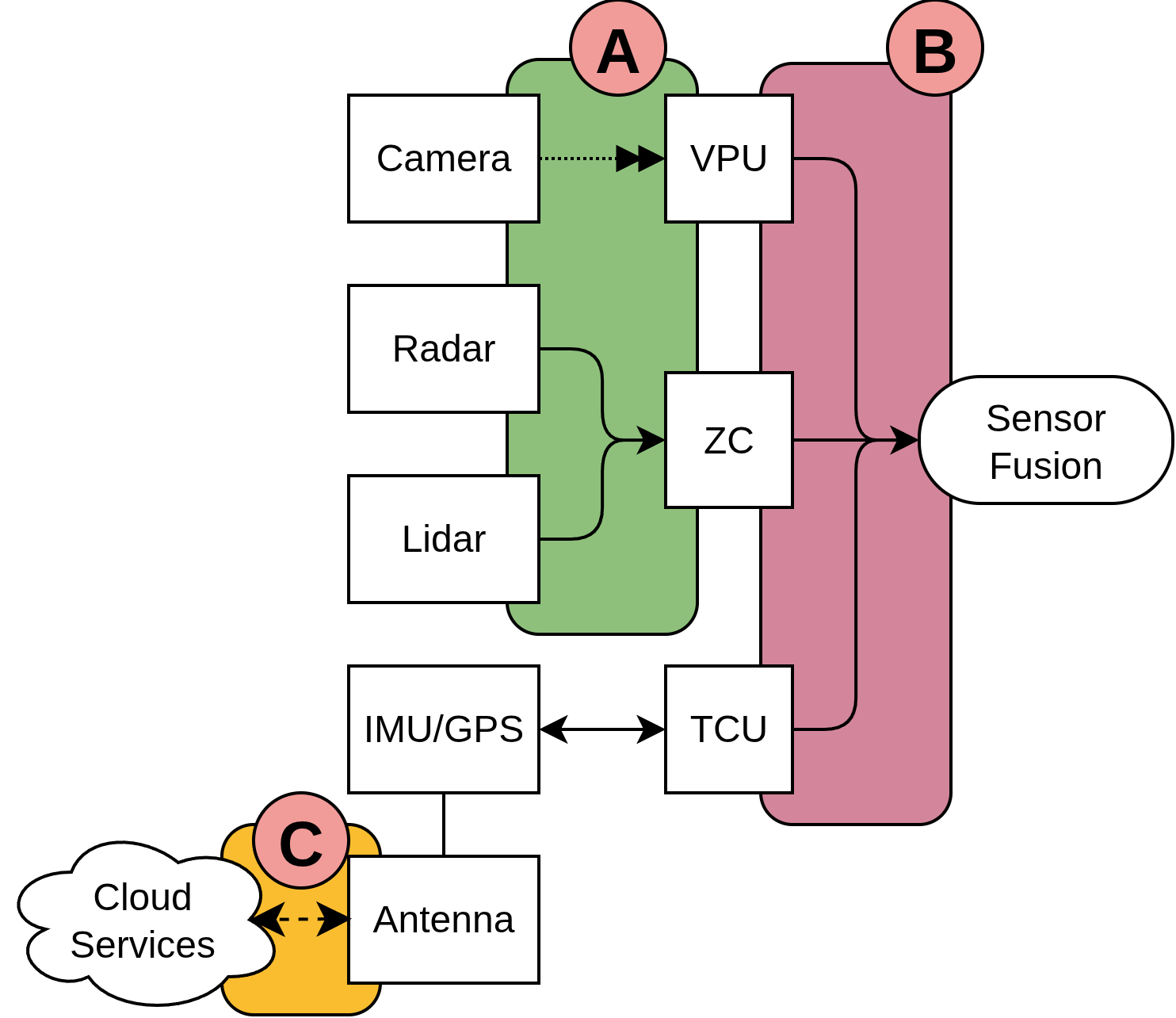}
            \Description{Potential sources of network timing variability in an SDV microservice workflow.}
            \caption{Potential sources of network timing variability in an SDV microservice workflow, including \textcircled{A} congestion caused by high throughput or bus saturation, \textcircled{B} protocol translation between heterogeneous service infrastructure, and \textcircled{C} transmission interference or network failure.}
            \label{fig:SRC_network}
        \end{figure}

        These delays may be significant in complex or high-traffic environments, such as SDVs. For example, higher sensor resolutions increase message size and transmission time, which in turn raises susceptibility to contention and interference \cite{choi2024EnhancingRecoveryPerformance, sperling2024ReducingCommunicationCost}. In legacy CAN-based architectures (a communication bus standard commonly used between ECUs), bandwidth limitations pressure infrastructure in ways that synchronization alone cannot solve, since synchronization mechanisms do not eliminate traffic saturation \cite{choi2024EnhancingRecoveryPerformance, sperling2024ReducingCommunicationCost}. Both of these scenarios can be seen in Figure \ref{fig:SRC_network} (\textcircled{A}), where sensor resolution may exceed bandwidth limitations depending on the data volume and wiring infrastructure. As congestion grows, less time remains for downstream computation, which tightens the coupling between network performance and the functional correctness of the platform.
        
        Traffic shaping and scheduled networking partially mitigate these effects. TSN and related deterministic Ethernet mechanisms tightly regulate transmission windows and can reduce contention under known traffic patterns \cite{do2024PerformanceAnalysisTraffic, teixeira2025DeterministicReliableSoftwareDefined}. However, when traffic patterns deviate from expected arrival curves, rebound effects can occur \cite{do2024PerformanceAnalysisTraffic}.
        
    \subsubsection{Routing}

        Communication latency depends, in part, on the delays introduced by routing and forwarding decisions across the network \cite{do2024PerformanceAnalysisTraffic, park2023AnalysisIPCANCommunication}. In heterogeneous deployments, different network components and paths may exhibit different performance characteristics, causing identical logical interactions to experience different timing behavior depending on where and how traffic is routed \cite{whaiduzzaman2021ResilientFogIoTFramework, sperling2024ReducingCommunicationCost}. For example, response times in a vehicle depend on whether a request is routed over Ethernet or a CAN bus.

        Several methods exist for managing timing variability from routing. Priority-based routing and scheduling can regulate contentious traffic and communication when combined with deterministic networking mechanisms \cite{teixeira2025DeterministicReliableSoftwareDefined, do2024PerformanceAnalysisTraffic}. Though, as noted in Section \ref{sssec:queuing}, traffic regulation can inadvertently result in queuing effects if traffic patterns are different than expected. The effects of routing decisions are often entangled with traffic load, queuing effects, and network interference, so routing decisions must be made with a clear understanding of system assumptions, infrastructure, and design.

    \subsubsection{Protocol Translation}

        Protocol translation and cross-layer interactions introduce timing variability whenever communication spans multiple networking technologies, middleware layers, or abstraction boundaries. In automotive and service-oriented systems, legacy CAN networks often coexist with Ethernet-based backbones, requiring gateways to translate between protocols while preserving timing constraints \cite{park2023AnalysisIPCANCommunication}. In Figure \ref{fig:SRC_network} (\textcircled{B}), the sensor fusion service must collate data received from different sources and potentially along different wiring types. These translations add processing overhead and introduce additional timing sensitivity at protocol boundaries, making end-to-end behavior harder to predict.

        The same issue appears at higher layers of the stack. Middleware frameworks such as DDS extend communication into the service layer through publish-subscribe abstractions and QoS settings, but these settings can themselves alter application timing. One example of this is reliability policies that may require acknowledgments before subsequent transmissions, directly affecting latency and throughput \cite{ferraro2023TimesensitiveAutonomousArchitectures}. Although such middleware improves coordination across ECUs and services, its timing behavior remains less well understood than lower-level communication mechanisms \cite{fraccaroli2023TimingPredictabilityIPbased}.
        
        More broadly, communication between abstraction layers in SDVs introduces nondeterminism: execution may span many layers, each using its own protocols and interfaces \cite{laclau2024DesignDynamicArchitectures, teixeira2025DeterministicReliableSoftwareDefined}. These abstractions improve modularity and manageability, but they also create additional interaction surfaces that affect timing behavior \cite{laclau2024DesignDynamicArchitectures, mauser2026MixedCriticalitySoftwareArchitectures}. Nondeterministic cross-layer coordination complicates timing analysis and may preclude the composability of local timing guarantees \cite{lee2008CyberPhysicalSystems, zhao2020RhythmComponentdistinguishableWorkload}.
        
    \subsubsection{Uncertainty}

        Network timing behavior is inherently uncertain when it depends on conditions that are only partially controlled or vary across time (for local uncertainty, see Sections \ref{sssec:scheduling} and \ref{sssec:synchronization}). Heterogeneous networks contain components with different performance characteristics, so communication timing cannot be assumed to be uniform across the system \cite{whaiduzzaman2021ResilientFogIoTFramework, sperling2024ReducingCommunicationCost}. For example, the network communication in Figure \ref{fig:SRC_network} (\textcircled{C}) is subject to interference and failure, and any disturbance in communication may impact downstream services. Even when deterministic networking mechanisms are present, timing remains sensitive to traffic composition, interference, and the specific management decisions applied to the network \cite{do2024PerformanceAnalysisTraffic, fraccaroli2023TimingPredictabilityIPbased}.

        Uncertainty is amplified in layered architectures, where execution spans boundaries between runtime environments and control layers \cite{laclau2024DesignDynamicArchitectures, teixeira2025DeterministicReliableSoftwareDefined}. In such settings, local controls may regulate specific parts of the communication path, but cannot fully eliminate the uncertainty caused by cross-layer interactions, environmental variation, or runtime contention \cite{camilli2018ZonebasedFormalSpecification, laclau2025EnhancingAutomotiveUser, zhao2020RhythmComponentdistinguishableWorkload}.
        
\subsection{Middleware Variability}

    \subsubsection{Scaling}\label{sssec:scaling}

        Deployment scaling is a key part of maintaining the performance of microservice-based systems. This is done by adding new microservice instances (horizontal scaling, common with container-based virtualization) or increasing the resources available to an instance (vertical scaling). When demand on a service increases, additional instances may be needed to prevent services from becoming overloaded. However, scaling services and adjusting traffic flows is not instantaneous, and is generally a reactive process \cite{almaruf2022UsingMicroserviceTelemetry}. In this sense, scaling decisions are a form of delayed feedback loop whose overall latency depends on both the architecture of the system and the responsiveness of the control mechanism.

        There are several sources of timing variability associated with scaling. First is the reaction speed of the controller. Autoscaling mechanisms typically rely on some combination of telemetry, thresholds, and historical data to decide when additional nodes or resources are required. Necessarily, any delay in the availability of system state (including from coarse sampling windows) degrades the reliability of the autoscaler. For example, a service may become overloaded before a scaling action is taken \cite{cheng2023ProScaleProactiveAutoscaling}. This delayed feedback can also exacerbate latency spikes, as traffic typically increases faster than new resources can be initialized \cite{gan2019anopensourcebenchmarksuite}. Additionally, under-provisioning or poorly distributing capacity can leave bottlenecks unresolved, so incorrect or partial actions also degrade autoscaler performance \cite{cheng2023ProScaleProactiveAutoscaling}.
        
        Second, reconfiguration actions have their own associated overhead. Scaling microservice instances often requires fetching container images, preparing the environment, and starting a new service replica. Depending on the virtualization environment, this may take seconds, while SLOs are often defined in milliseconds \cite{wen2024VirtualizationMicroserviceArchitecture, luo2022PowerPredictionMicroservice}. If additional virtual machines are deployed, horizontal scaling can be even slower \cite{wan2018ApplicationDeploymentUsing}. Fine-grained resource management operations also take time, which imposes a lower bound on the speed with which controllers can take mitigating action. Latency spikes shorter than this response window cannot be corrected within the mitigation deadline, and may instead lead to harmful actions or oscillations between scaling operations \cite{qiu2020FIRMIntelligentFinegrained}.
        
        Finally, scaling perturbs the behavior of a system when the action changes state or reconfigures nodes where work is being executed. Stateful migrations can take several milliseconds, and introduce variability during the transition \cite{delimitrou2014QuasarResourceefficientQoSaware}. Similarly, if a transient overload disappears after additional replicas are deployed, the system state may differ from the fault-free state assumed by monitoring tools \cite{cornacchia2026ObservabilityEatingYour}. Vertical scaling introduces related effects, as adjusting resource capacity can add latency during the scaling process \cite{wan2018ApplicationDeploymentUsing}.
        
        Deployment scaling creates a trade-off in microservice timing behavior. Without scaling, overloaded services may accumulate queues and increase response times, but scaling mechanisms add feedback delays and overhead that are difficult to measure or predict. Because scaling decisions form a feedback loop between observability, execution, and control layers (see Figure \ref{fig:BCK_control_layers}), their effects on timing behavior are difficult to categorize. While autoscaling is studied extensively in cloud environments and self-adaptive systems, it is often absent or externalized from execution and timing models \cite{arcaini2015ModelingAnalyzingMAPEK, xie2026SlowpokeThroughputOptimization, zeng2025CASLOJointScaling}.

    \subsubsection{Node Failure}

        Individual node failures are a common occurrence in large microservice deployments, but rarely appear as a clean stop. Microservice instances may crash, dependencies may time out, or configuration changes may produce cascading errors. Because node failures have timing implications beyond the failure itself, we include them here, as a coordination-level source of variability. In addition to the usual delays between detection and corrective action (see Section \ref{sssec:scaling}), a failure often first manifests as a slow or unresponsive service. This forces operations to proceed, possibly without progress being made on any connected service \cite{barroso2019DatacenterComputerDesigning, mohammad2025ResilientMicroservicesSystematic}. From a timing perspective, this means that failure results in both a loss of capacity and long and unpredictable response times.

        These effects have an outsized impact on tail latencies. While temporary periods of high latency may have a negligible effect on smaller systems, they can dominate the overall performance of larger systems \cite{dean2013TailScale}. Since a small number of slow components can affect many end-to-end requests, node failures contribute to timing variability both directly (by delaying or preventing responses) and indirectly (by increasing retry traffic and load on healthy services).
        
        Failure handling also reintroduces many of the timing effects presented in Section \ref{sssec:scaling}. For example, an orchestrator may need to restart an unresponsive container \cite{barroso2019DatacenterComputerDesigning, rodrigues2005RobustServicesDynamic}. Before corrective action is taken, requests may be incorrectly routed to unavailable services, which can result in timeouts or retry loops. Any microservices attempting to communicate with the failed service may also fail during this time, causing the failure to propagate through a system \cite{sampaio2019ImprovingMicroservicebasedApplications, liu2021ApproachModelingAnalyzing}.

    \subsubsection{Mixed-Criticality Interference}

        Modern SDV platforms increasingly consolidate heterogeneous workloads with fundamentally different timing and safety requirements into shared hardware and software environments \cite{mauser2026MixedCriticalitySoftwareArchitectures, sperling2024ReducingCommunicationCost}. HPCs may combine microcontroller and microprocessor functionality within a single System-on-Chip; safety-critical control functions and less critical service-oriented applications often coexist within the same execution platform. For example, workflows for collision avoidance and turn signal activation may share much of the same infrastructure, but have vastly different priorities and timing requirements. This creates a mixed-criticality environment where applications differ in the strictness of their timing constraints and levels of acceptable interference \cite{mauser2026MixedCriticalitySoftwareArchitectures}.

        At the hardware level, features that improve flexibility and performance, such as caches, memory hierarchies, and speculative execution, also introduce variability that complicates timing analysis \cite{camilli2018ZonebasedFormalSpecification, lee2008CyberPhysicalSystems}. Hardware partitioning and static core allocation can improve predictability by isolating critical workloads, but these measures reduce flexibility and efficiency. Safety-critical embedded systems have historically managed this tradeoff by freezing hardware configurations in order to preserve validated timing guarantees and avoid recertification costs \cite{lee2008CyberPhysicalSystems, mauser2026MixedCriticalitySoftwareArchitectures}.
        
        \begin{figure}
            \centering
            \includegraphics[width=0.5\linewidth]{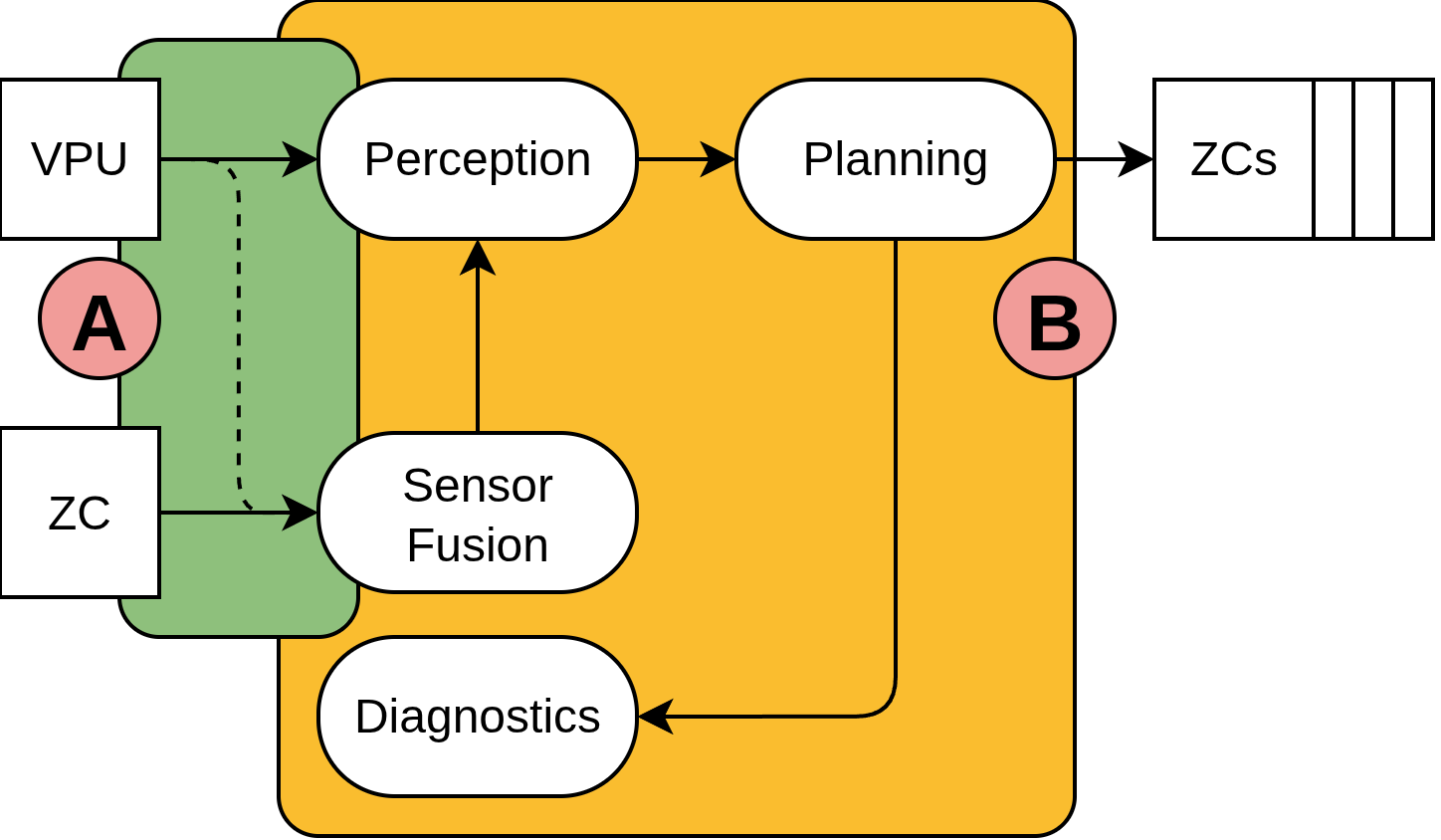}
            \Description{Potential sources of timing variability from coordination and control mechanisms in an SDV microservice workflow.}
            \caption{Potential sources of timing variability from coordination and control mechanisms in an SDV microservice workflow, including \textcircled{A} mixed-criticality workloads and \textcircled{B} microservices subject to scaling, reconfiguration, and failure.}
            \label{fig:SRC_coordination}
        \end{figure}
        
        In distributed microservice-based systems, mixed-criticality interference extends to network effects, such as communication delays, dependency interactions, asynchronous execution, and congestion \cite{kamboj2025LeveragingPetriNets}. Cloud and edge platforms often separate latency-critical and best-effort workloads through orchestration policies designed to avoid SLO violations, and automotive systems require analogous isolation mechanisms at the network level, such as VLAN segmentation, bandwidth reservation, and TSN-based scheduling \cite{laclau2024DesignDynamicArchitectures, laclau2025EnhancingAutomotiveUser, zhao2020RhythmComponentdistinguishableWorkload, ferraro2023TimesensitiveAutonomousArchitectures, fraccaroli2023TimingPredictabilityIPbased}. The collision avoidance system in Figure \ref{fig:SRC_coordination} (\textcircled{A}) must manage mixed-criticality interference if the vision system creates a priority workflow that bypasses other parts of the sensor package. Without resource-aware orchestration, such systems must provision for worst-case demand. This increases overhead and reduces overall efficiency \cite{laclau2024DesignDynamicArchitectures}. Because mixed-criticality interference appears across multiple layers of a distributed system, it presents a problem of system-level coordination, rather than purely local scheduling.

    \subsubsection{Coordination and Reconfiguration Overhead}\label{sssec:coordination_and_reconfiguration}

        Microservice-based SDV platforms rely on orchestration mechanisms to manage application lifecycles and resource allocation within virtualized environments \cite{pan2024SoftwareDefinedVehiclesModelBased}. These orchestrators typically assume that applications are mapped to execution contexts in advance, with resources reserved for safety-critical workloads according to QoS requirements. The remaining resources are then allocated dynamically in response to runtime conditions through resource-aware control logic \cite{laclau2025EnhancingAutomotiveUser, wan2018ApplicationDeploymentUsing}. As a result, execution timing is closely tied to the responsiveness of orchestration mechanisms and the speed at which they can adjust and distribute deployment configurations.

        In SDVs, orchestration decisions interact continuously with physical processes and environmental inputs, forming closed feedback loops between sensors, software, and actuators \cite{lee2008CyberPhysicalSystems}. Changes in the sensed state may trigger application launches, shutdowns, or mode switches, and some architectures split these responsibilities between offboard optimization engines and onboard heuristic schedulers that combine precomputed configurations with runtime estimates \cite{laclau2024DesignDynamicArchitectures, laclau2025EnhancingAutomotiveUser, dustdar2023DistributedComputingContinuum, whaiduzzaman2021ResilientFogIoTFramework}. This decomposition can improve responsiveness and scalability, but it also increases the coordination required to maintain a consistent view across controllers and execution environments.
        
        When conditions change, the system must re-optimize its runtime mode and propagate a new configuration. These transitions are not instantaneous. Service discovery, optimization, and configuration distribution all contribute additional overhead, and each mode transition must complete within explicit time bounds in order to preserve a consistent system state \cite{camilli2018ZonebasedFormalSpecification, fraccaroli2023TimingPredictabilityIPbased, laclau2025EnhancingAutomotiveUser}. More fundamentally, this approach assumes that such a consistent global system state is continuously available. When that state is incomplete or unavailable, the system may revert to default behavior or make less effective coordination decisions \cite{laclau2025EnhancingAutomotiveUser}. Coordination overhead is therefore itself a source of timing variability: its efficiency is partially determined by its ability to collect state information, compute a new configuration, and distribute that information in a timely manner.

    \subsubsection{Observer Effects}

        Observability mechanisms collect traces, metrics, and event logs that make it possible to analyze execution behavior that is not directly visible in distributed microservice-based systems \cite{cornacchia2026ObservabilityEatingYour, vanderaalst2012ProcessMiningOverview}. Because individual services typically expose only local request/response information, end-to-end behavior must often be reconstructed by aggregating fragmented telemetry across abstraction layers and components \cite{gu2023TrinityRCLMultiGranularCodeLevel, ashok2024TraceWeaverDistributedRequest, otero2024LightweightDistributedTelemetry, otero2025ExtensibleLightweightFramework}. Runtime reconstruction of execution graphs helps recover dependency relations, identify bottlenecks and failure propagation paths, and supports diagnosis and resource management in systems whose declared architecture may differ from their actual runtime behavior \cite{almaruf2022UsingMicroserviceTelemetry, wang2020WorkflowAwareAutomaticFault, yu2021MicroRankEndtoEndLatency, kamboj2025LeveragingPetriNets, lin2018MicroscopePinpointPerformance, soldani2023AnomalyDetectionFailure}.

        However, this reconstruction process is itself computationally difficult. In asynchronous and highly parallel environments, associating parent and child requests requires reasoning over large numbers of possible causal relationships, and the number of candidate correlations can grow combinatorially with the degree of concurrency \cite{ashok2024TraceWeaverDistributedRequest, kamboj2025LeveragingPetriNets, liu2021MicroHECLHighEfficientRoot}. When service meshes are used, as in some cloud, edge, or SDV deployments, the overhead of maintaining a sufficiently detailed runtime view can make real-time reconstruction infeasible \cite{cornacchia2026ObservabilityEatingYour, gu2023TrinityRCLMultiGranularCodeLevel}. Noisy telemetry, incomplete logs, coarse sampling, and dynamically changing baselines can all distort the recovered execution graph and weaken diagnosis accuracy \cite{vanderaalst2012ProcessMiningOverview, yu2021MicroRankEndtoEndLatency, cornacchia2026ObservabilityEatingYour, soldani2023AnomalyDetectionFailure}. Accuracy is crucial, especially when profiling tail latency or short-lived anomalies, as reconstruction errors can directly impact performance attribution.
        
\begin{figure}[h]
            \centering
            \includegraphics[width=0.9\linewidth]{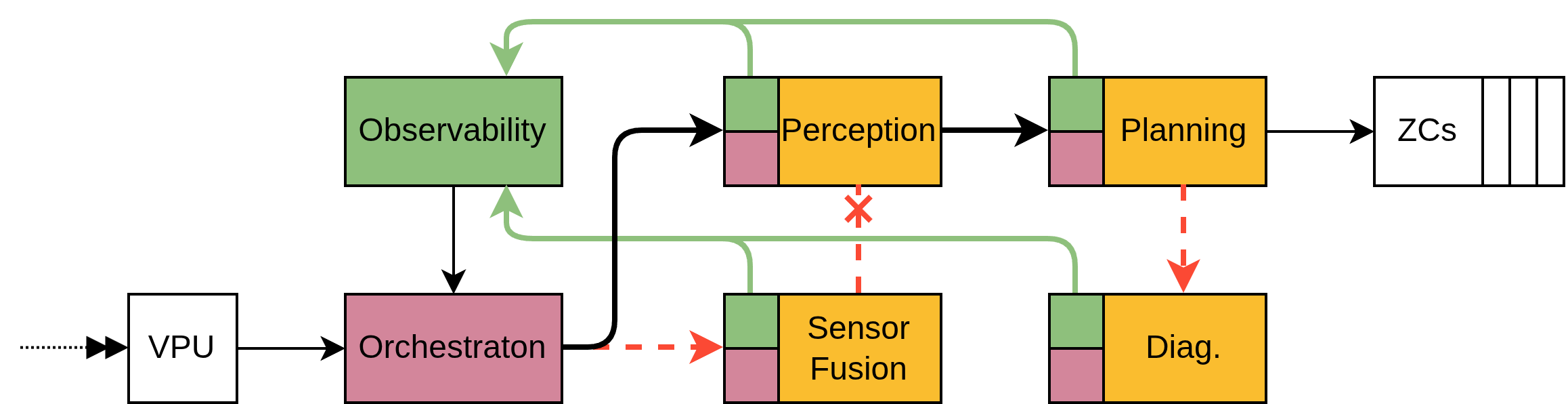}
            \Description{An example microservice workflow, including orchestration and observability layers. When priority flows are enforced, several directed paths are removed from the graph.}
            \caption{When a priority workflow is enforced between the VPU and the perception service, traffic flows to and from the sensor fusion service are impacted. Additional overhead from the observability controller may further restrict the resources of lower-priority services.}
            \label{fig:SRC_observer}
        \end{figure}

        More importantly, monitoring is not passive. Observability layers compete with application services for CPU, memory, and network resources, increase startup and response-time overhead, and in extreme cases can substantially amplify tail latency \cite{ashok2024TraceWeaverDistributedRequest, cornacchia2026ObservabilityEatingYour, wang2020WorkflowAwareAutomaticFault, dean2013TailScale}. Instrumentation may also change the structure of the system itself: context propagation can require invasive code changes, telemetry collection services are often inserted into request paths as middleware (see Figure \ref{fig:SRC_observer}), and some frameworks actively modify concurrent event streams in order to make execution easier to analyze \cite{otero2024LightweightDistributedTelemetry, otero2025ExtensibleLightweightFramework, ashok2024TraceWeaverDistributedRequest}. As a result, observability tools alter traffic patterns, contention, and end-to-end latency, and may even change the effective execution graph being measured \cite{soldani2023AnomalyDetectionFailure, wen2024VirtualizationMicroserviceArchitecture}. Sampling is commonly used to mitigate this overhead, but this creates a direct tradeoff between measurement fidelity and system perturbation: reducing telemetry volume improves scalability, but longer sampling windows risk missing rare or short-duration events \cite{cornacchia2026ObservabilityEatingYour, otero2024LightweightDistributedTelemetry}.

\section{Timing Propagation}\label{sec:timing_propagation}

    Timing variability in a single component of a microservice-based system rarely remains isolated, and is instead able to propagate across service boundaries and affect the system as a whole. This section examines the structural mechanisms through which timing variability spreads and interacts.

\subsection{Dependency Graphs}\label{ssec:dependency_graphs}

\begin{figure}
        \centering
        \includegraphics[width=0.6\linewidth]{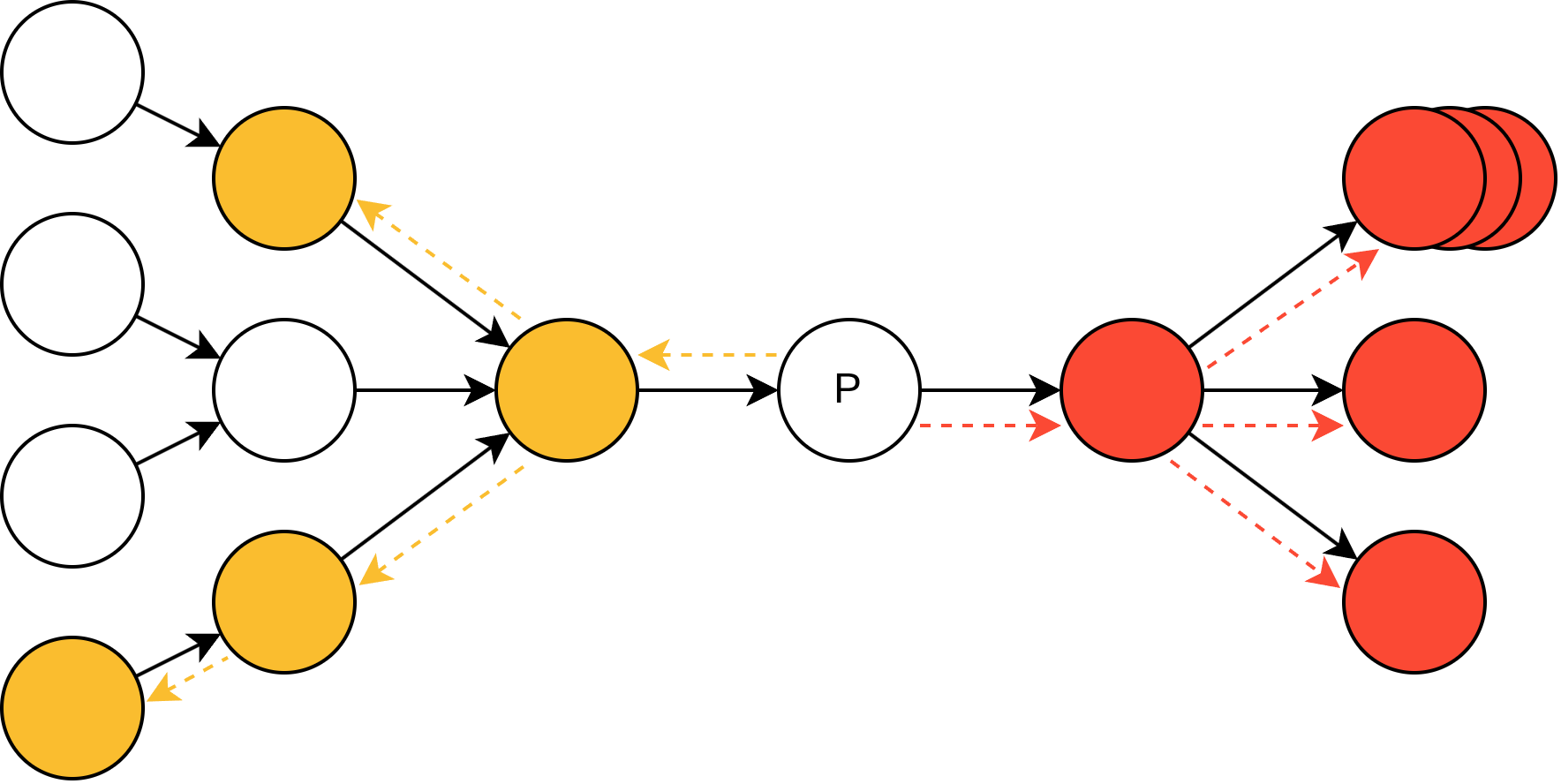}
        \Description{An abstracted dependency graph showing directional links between services. All downstream services are highlighted red, and half of the upstream service are colored yellow; these represent a slowdown resulting from a single service which has not been localized.}
        \caption{An increase in execution time of service P can impact both upstream (yellow) and downstream (red) services. The propagation is based on the dependency relations of individual requests, so the declared execution graph may be insufficient for determining the location or scale of the effects.}
        \label{fig:PRP_dependency}
    \end{figure}

    It is rare for a request to be handled by a single component in a microservice application; instead, requests traverse a series of dependent services to accomplish some task \cite{ashok2024TraceWeaverDistributedRequest}. Dependency graphs are a common way of modeling these invocation chains, and are a powerful abstraction for analyzing how timing effects propagate through a system. Any timing variability introduced by one service or network component may affect every other component which depends on it, though this is often unpredictable and dependent on how services interact.  Figure \ref{fig:PRP_dependency} shows how a slowdown in one service (P) can affect many other connected services. However, a service may be unaffected if, for example, it has a sufficient timing buffer to absorb the impact.

    The topology of the dependency graph determines how timing effects compose and spread. In serial execution paths, latency accumulates across dependent services; in concurrent paths, a slow branch may stall aggregation and delay the final response; and shared dependencies can cause contention or queuing delays to travel upstream to otherwise independent services. Similar local delays will impact the system differently depending on their location, degree of connection, and distance from the critical path \cite{hu2025LSRAMLightweightAutoscaling}.
    
    Understanding the topological impacts on timing behavior is crucial for most timing models. Because many SLOs are decomposed into per-service objectives, graph structure determines how timing budgets should be allocated \cite{hu2025LSRAMLightweightAutoscaling, kannan2019GrandSLAmGuaranteeingSLAs}. Similarly, dependency graphs are used to infer which services are responsible for performance anomalies, because the observed symptoms may appear far from the cause \cite{wu2020MicroRCARootCause, kannan2019GrandSLAmGuaranteeingSLAs}. In both cases, the underlying model must identify the constituent structures of the dependency graph and estimate their contribution to the behavior of the system.
    
    The scale of microservice applications can make identifying these propagation structures difficult. Many applications contain hundreds of loosely-coupled services, and studies have shown that microservice call graphs can differ substantially from simplified DAG representations \cite{gan2019anopensourcebenchmarksuite, luo2022InDepthStudyMicroservice}. As the graph grows, so does the number of points where timing variability can unexpectedly enter the system \cite{ashok2024TraceWeaverDistributedRequest, kamboj2025LeveragingPetriNets, liu2021MicroHECLHighEfficientRoot}, and misattributing even a single dependency can have a dramatic impact on tail latency \cite{gan2019anopensourcebenchmarksuite}. Because runtime graphs often diverge from their design-time counterparts, accurate reproduction is crucial to predicting and measuring the timing behavior of microservice-based systems.

\subsection{Resource Sharing}\label{ssec:resource_sharing}

    When multiple services interact, either from within the same execution environment or across a network, their resource demands can interact in ways that are difficult to predict. Services running on the same machine compete for CPU or memory, and may compete with distributed services for global resources, such as network bandwidth or access to shared file systems \cite{dean2013TailScale}. As utilization increases, the performance penalties from resource contention become more significant, as each additional request consumes capacity that may be needed elsewhere \cite{barroso2019DatacenterComputerDesigning}. In this way, shared resources create implicit timing dependencies between services that may not directly invoke each other.

\begin{figure}
        \centering
        \includegraphics[width=0.6\linewidth]{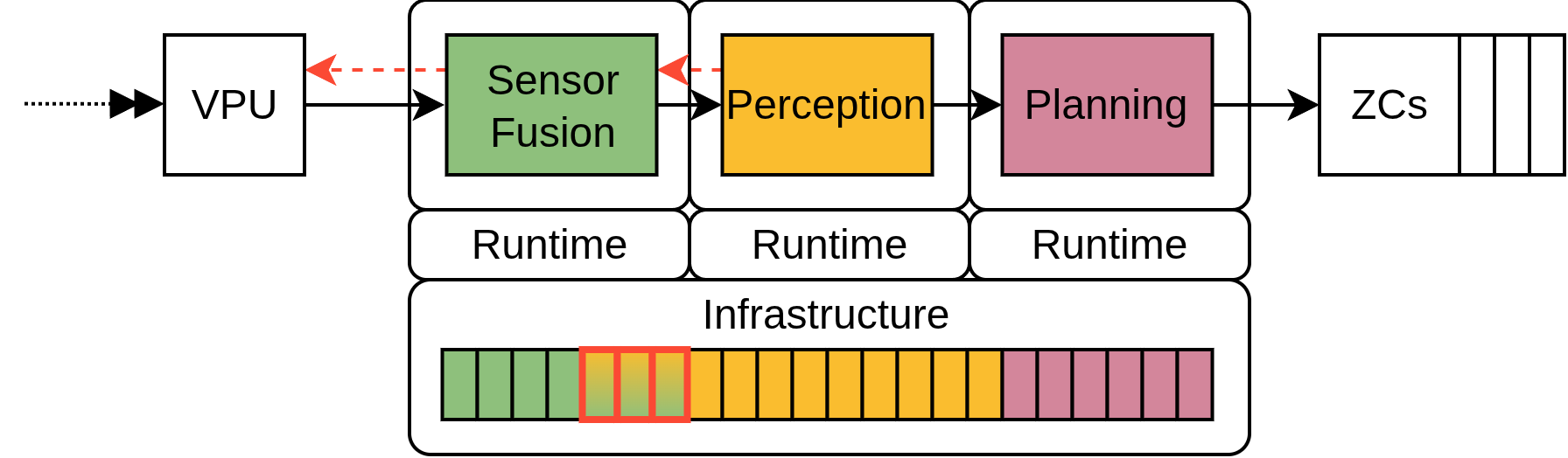}
        \Description{The Sensor Fusion and Perception services interfere with each other because their resources come into conflict. Here, the Perception service uses too much memory, which causes Sensor Fusion to perform erratically.}
        \caption{Contested resources can directly impact the performance of co-located services. If the perception service occupies too much memory or CPU time, the sensor fusion service may behave erratically, propagating timing effects to its dependencies.}
        \label{fig:PRP_contention}
    \end{figure}

    The impact of these implicit dependency relations is especially relevant in microservice applications, where resource sharing can occur across different abstraction levels. Within a single virtualization host, co-located microservices may collectively demand more resources than are available, causing unexpected or undefined behavior (as shown in Figure \ref{fig:PRP_contention}) \cite{sampaio2019ImprovingMicroservicebasedApplications}. This can cause interference within the same node even when services do not directly communicate with each other \cite{lin2018MicroscopePinpointPerformance}. If a service experiencing contention communicates across a network, it can lead to bottlenecks and SLO violations \cite{kannan2019GrandSLAmGuaranteeingSLAs}. Because shared communication paths can also cause timing variability to spread through a system (e.g., via queuing, link saturation, etc.), we consider network congestion to be a form of resource contention.
    
    The impacts of contention depend on which resource is saturated, which workloads share it, and how tightly they are coupled. In general, this means that CPU utilization has the strongest impact on response times, but the resulting delays are not uniform \cite{luo2022InDepthStudyMicroservice}. However, because the propagation of timing effects by shared resources is determined more by infrastructure capacity than by logical dependency relations, it can be difficult to find the source of performance bottlenecks in large or complex systems \cite{harchol-balter2003SizebasedSchedulingImprove, wang2024AutothrottlePracticalBiLevel}.
    
    \begin{figure}
        \centering
        \includegraphics[width=\linewidth]{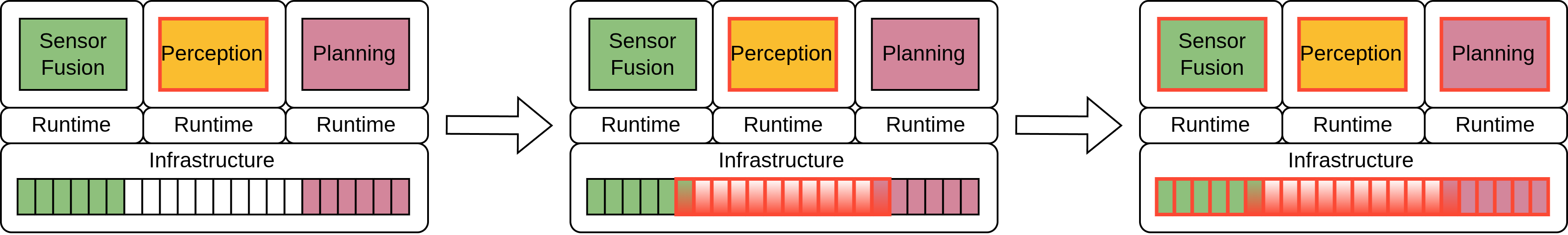}
        \Description{Three panels show the progression of thrashing behavior. First, the Perception service goes down. Then, the startup overhead for the Perception service conflicts with the resource requirements for Sensor Fusion and Planning. Finally, Sensor Fusion and Planning show signs of overload.}
        \caption{A detected failure can lead to an overcommitment of resources (CPU, memory, or network capacity). The resulting overload can be misinterpreted as additional failures.}
        \label{fig:PRP_thrashing}
    \end{figure}
    
\subsection{Feedback and Control Loops}\label{ssec:feedback_and_control_loops}

    Coordination and control mechanisms work with observability tools to adapt to runtime conditions and improve system performance. This results in feedback loops between the observation of system state and changes to the conditions under which future requests are executed. The feedback loops discussed in this section typically appear in the interactions between middleware components, though, in principle, they can also include other control mechanisms, such as schedulers or network controllers. Control mechanisms may be centralized or decentralized, but microservice-based systems typically rely on centralized controllers to manage configuration changes \cite{delemos2013SoftwareEngineeringSelfAdaptive}.

    Control structures can create timing dependencies between components that would otherwise be independent, particularly when controllers share information, form observation and control loops, or otherwise coordinate their actions \cite{camilli2018HighlevelPetriNetbased}. For example, a measurement delay can postpone corrective action, and a poorly coordinated configuration change can move resource contention from one part of the system to another \cite{barroso2019DatacenterComputerDesigning}.
    
    The speed and effectiveness with which controllers are able to detect and react to problems influences whether those problems are mitigated, ignored, or amplified. A controller that misinterprets symptoms or reacts to outdated information can produce harmful effects. Rodriques et al. describe a scenario where an attempt to recover from a host failure causes network traffic that is interpreted as additional host failures, creating a positive feedback loop (see Figure \ref{fig:PRP_thrashing}) \cite{rodrigues2005RobustServicesDynamic}. This example shows how a local disturbance can be propagated directly by control layers: the initial failure changes the system state, the response creates resource contention, and the resulting queues are interpreted as cascading failures. In such cases, the mechanisms intended to restore normal operation may instead recursively respond to and spread timing variability.
    
\subsection{Amplification Effects}\label{ssec:amplification_effects}

    In addition to the structural mechanisms discussed earlier, there are also transformative propagation mechanisms that magnify local timing variabilities into larger system-level effects. These mechanisms relate to the structural interdependence of system components (e.g., utilization, fan-out degree, or distance from the critical path), and cause nonlinear timing behaviors in the system.
    
    \subsubsection{Tail Amplification}

        When multiple services interact, the timing variability of individual components may become magnified at the system level \cite{dean2013TailScale}. In microservice-based systems, a request may traverse many services with interacting dependencies and resource requirements, and even a small fraction of long latencies can negatively impact the perceived responsiveness of the system. Barroso et al. illustrate this effect with a request that must collect responses from 100 services in parallel: even if each service individually has only a small probability of responding slowly, the probability that at least one response delays the full response is large enough that most users will experience a delay \cite{barroso2019DatacenterComputerDesigning}.

        \begin{figure}
            \centering
            \includegraphics[width=0.7\linewidth]{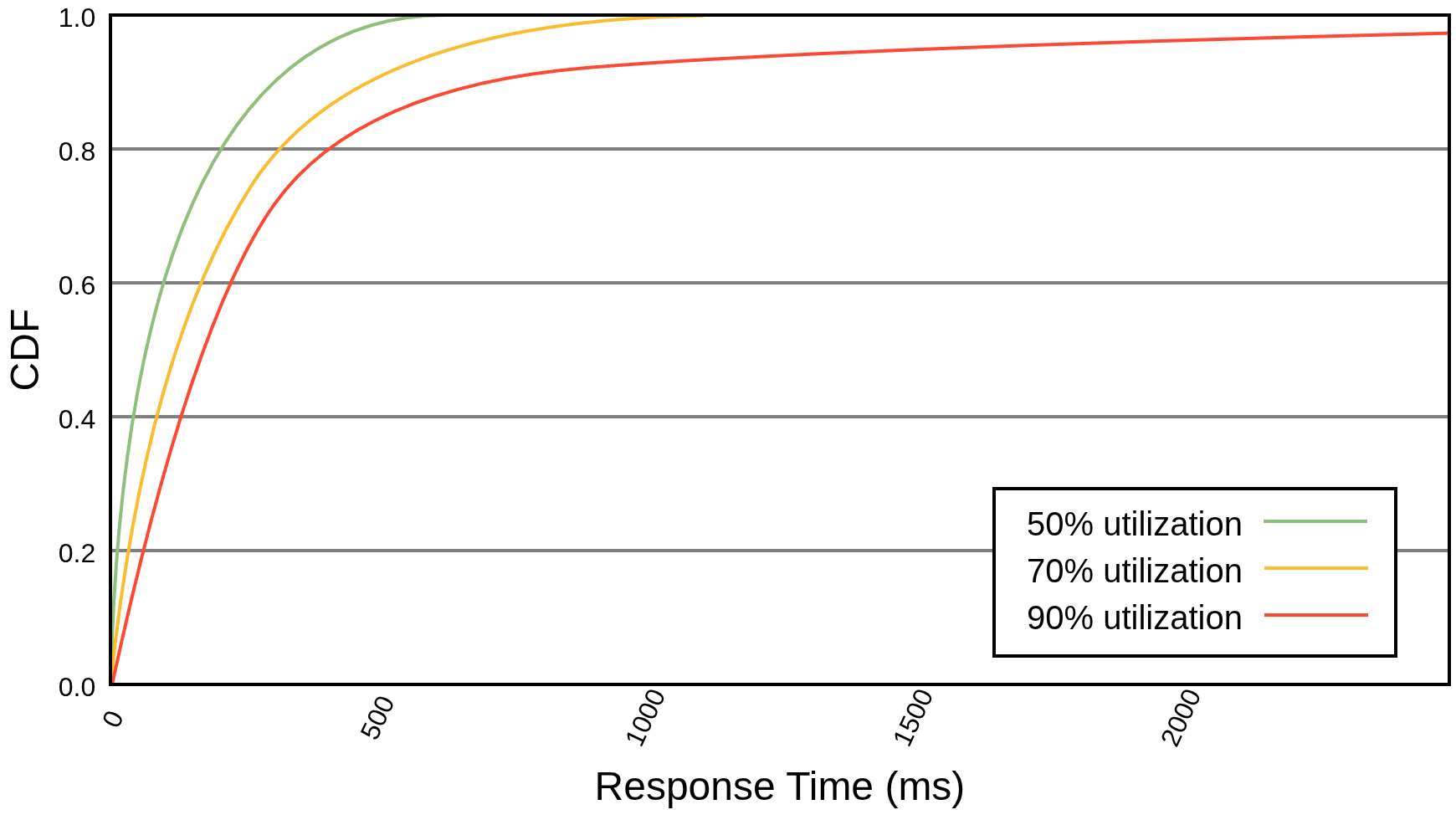}
            \Description{Three latency curves have progressively longer tails as utilization increases from 50\% to 90\%.}
            \caption{Tail latency is amplified by increased processor utilization \cite{li2014TailsOfTheTail}.}
            \label{fig:PRP_tail_amplification}
        \end{figure}
        
        Tail amplification is closely tied to the execution graph structure. Microservice applications may contain many individual services, and studies show that traffic patterns often follow long-tail latency distributions \cite{luo2022InDepthStudyMicroservice}. Large and highly-connected execution graphs provide more opportunities for a request to be delayed by a connected component. Tail amplification is more pronounced in microservice applications than in monoliths, because the end-to-end latency of a request depends on the combined behavior of many independently-variable service latencies \cite{gan2019anopensourcebenchmarksuite}, and this problem is exacerbated in systems with many parallel execution paths or with high system overhead \cite{barroso2019DatacenterComputerDesigning}. Figure \ref{fig:PRP_tail_amplification} shows the impact of processor utilization on tail latencies in an edge testbed environment \cite{li2014TailsOfTheTail}, though tail amplification effects are general to other microservice deployments, including SDVs.

        Because the performance of coordination mechanisms can degrade when latency distributions become long-tailed, control layers are also susceptible to tail amplification \cite{bailis2014CoordinationAvoidanceDatabase}. A single slow response can inflate the average latency of an entire trace, which may lead to incorrect control actions if the observability tool is not able to identify the root cause \cite{ashok2024TraceWeaverDistributedRequest}. In a sense, tail amplification is a propagation effect caused by unexpected interactions between the graph structure and component-level timing variability.
        
    \subsubsection{Anomaly Amplification}

        Local problems or component failures can cascade into larger downstream timing anomalies (i.e., an undefined state, instability, or unexpected behavior). For example, a buffer overflow may cause a service to behave erratically. An anomaly in one service can propagate and become much larger than the scale of the original problem would suggest. In some cases, this can negatively impact the availability of the full application \cite{liu2021MicroHECLHighEfficientRoot}. As with tail amplification, anomaly amplification appears as a set of system-level symptoms that can be difficult to isolate and trace back to their source \cite{otero2024LightweightDistributedTelemetry}.

        The service chain is the most common propagation path for this type of error amplification. If one service becomes slow or unavailable, upstream services may form queues or begin consuming additional resources. In one example, a memory leak in a payment service indirectly causes a backlog and CPU pressure on the upstream order service \cite{cornacchia2026ObservabilityEatingYour}. These cascading failures can be multiplicative in synchronous execution paths, where services are tightly coupled \cite{wang2023ComplexBehavioralInteraction}. Queuing and QoS violations caused by amplified anomalies can spread across application boundaries (such as from databases to front-end applications), effectively turning a degradation in one service into resource pressure elsewhere \cite{gan2019anopensourcebenchmarksuite, soldani2023AnomalyDetectionFailure, yu2021MicroRankEndtoEndLatency}.
        
        Though anomalies typically cascade through dependency relations, they are not limited to direct invocations. Co-located microservices can propagate failures when the failure affects the availability of shared resources \cite{gu2023TrinityRCLMultiGranularCodeLevel}. This makes propagation pathways difficult to predict from the dependency graph alone, especially when systems are large or distributed across many execution contexts \cite{dustdar2023DistributedComputingContinuum, wang2022ResearchCurrentSituation}.
        
        Because amplification generally produces many symptoms from one cause, diagnosing and localizing the source can be difficult. As such, anomaly detection frameworks often try to model propagation opportunities explicitly \cite{wu2020MicroRCARootCause, lin2018MicroscopePinpointPerformance}. However, because misjudging the number or location of faulty components can lead to other performance problems (see Section \ref{ssec:feedback_and_control_loops}), it is generally simpler to interrupt anomaly cascades than to prevent them \cite{liu2022ModellingAnalysingReliability, dinh-tuan2020DevelopmentFrameworksMicroservicebased}.
        
        \begin{figure}
            \centering
            \includegraphics[width=0.8\linewidth]{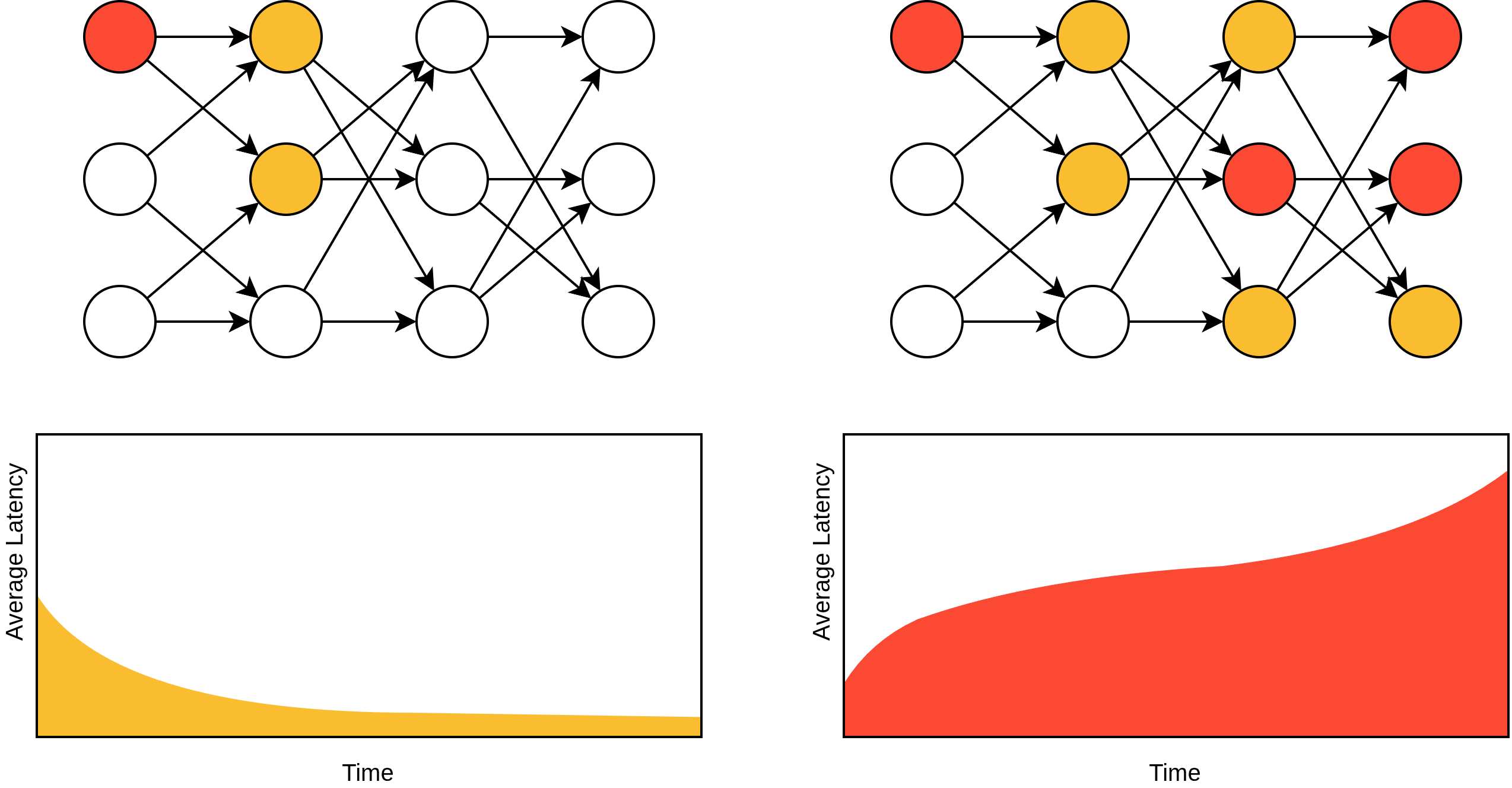}
            \Description{On the left, message queues are able to clear quickly, and average application latency remains manageable. On the right, accumulating message queues lead to increased average latency and several overloaded services.}
            \caption{If services are not able to quickly clear their message queues, the accumulated delay can spread to their dependencies, increasing application latency \cite{saramaki2024CriticalDelayAccumulation}.}
            \label{fig:PRP_delay_accumulation}
        \end{figure}

    \subsubsection{Delay Accumulation}

        In microservice-based systems, end-to-end latency is the accumulation of local execution steps and delays (including coordination time), which are distributed across the execution path \cite{fraccaroli2023TimingPredictabilityIPbased}. Though the individual sources of local delay are generally well known, in practice, each component acts as a black box that obscures the contributions of each stage to overall latency \cite{leboudec2001NetworkCalculus, sperling2024ReducingCommunicationCost}. As an example, small changes to hardware configuration can affect how the system responds to increased load; many such changes go undocumented \cite{laclau2024DesignDynamicArchitectures}. In general, implementation details restrict how much can be known about a system that hasn't been audited and frozen \cite{fraccaroli2023TimingPredictabilityIPbased, sperling2024ReducingCommunicationCost}. Consequently, classical task-level metrics such as maximum lateness and completion time are only useful within specific execution contexts, and do not generalize in a straightforward way to system-level timing behavior \cite{leboudec2001NetworkCalculus, stankovic1995ImplicationsClassicalScheduling}.
        
        The accumulative effects of local timing behavior become more pronounced when workloads are concurrent or include resource contention \cite{camilli2021FormalSpecificationVerification, fan2013ModelingOptimizingResource}. In these systems, contention is expressed as queuing effects, and mitigation mechanisms may themselves conflict with the timing requirements of individual components \cite{camilli2018FormalFrameworkSpecifying, fan2013ModelingOptimizingResource, sperling2024ReducingCommunicationCost}. If a system is not able to clear queues quickly enough, they can have a multiplicative effect that can quickly spread beyond the original source of the delay (see Figure \ref{fig:PRP_delay_accumulation}). This behavior can form amplifying feedback loops when components are composed together. End-to-end latency becomes an emergent result of the aggregated delays across all execution stages, which is affected by architecture, local execution behavior, and control policies \cite{kannan2019GrandSLAmGuaranteeingSLAs, teixeira2025DeterministicReliableSoftwareDefined}. Predicting the timing behavior of these systems depends largely on how scheduling and resource-management policies arbitrate among components and execution environments \cite{stankovic1995ImplicationsClassicalScheduling}.
        
        For distributed real-time systems (including some SDV subsystems), ``correctness'' also depends on the timely coordination of discrete events \cite{camilli2018ZonebasedFormalSpecification}. A safety-critical component may need to respond within a strict deadline, and incorrect event ordering can interfere with normal operation. Meeting SLOs requires satisfying timing requirements across the full execution path, regardless of the complexity of local analysis (e.g., for determining interference or control-loop interactions) \cite{arcaini2015ModelingAnalyzingMAPEK, kannan2019GrandSLAmGuaranteeingSLAs}. Because shared-resource scheduling is computationally difficult, formal models often must estimate the interactions between managed and unmanaged components to accurately predict timing behavior \cite{camilli2021FormalSpecificationVerification, stankovic1995ImplicationsClassicalScheduling, camilli2018ZonebasedFormalSpecification}. More generally, distributed timing analysis must account for dynamic resource reallocation, as the majority of black-box delay accumulation stems from adaptive behaviors at runtime, rather than from the design-time (a priori) structure of the system \cite{fan2016FormalAspectOrientedMethod, whaiduzzaman2021ResilientFogIoTFramework}.

\section{Impacts of Propagation}\label{sec:observable_timing_behavior}

    This section examines the observable system-level behaviors resulting from timing propagation and amplification. Because some effects can have multiple causes, we will focus on which behavioral patterns emerge, how they are observed, and their impacts on the overall health and stability of the system.

\subsection{Tail-Dominant Behavior}

    Because microservice requests tend to depend on many interconnected services, timing propagation in large systems often produces tail-dominant behavior. High-latency events that may have a negligible impact in smaller systems can dominate the perceived responsiveness of larger systems. This skewing of end-to-end latency distributions is largely dependent on the number and complexity of the interactions in a system, and becomes more pronounced when concurrency, network requests, and queuing are present \cite{barroso2019DatacenterComputerDesigning, dean2013TailScale}.

    In systems with large dependency graphs, a single overloaded (or poorly configured) service can degrade end-to-end latency by several orders of magnitude \cite{gan2019anopensourcebenchmarksuite}. Shared dependencies, especially those relying on cloud infrastructure, exacerbate contention or failure effects significantly, and can produce tail latencies with extremely large variance. In some cases, latency fluctuations can exceed the median response time by hundreds of times \cite{zhao2020RhythmComponentdistinguishableWorkload}. Average latency is unrepresentative of the user experience in such cases.

\subsection{Queue Propagation}\label{ssec:queue_propagation}

    A localized increase in latency frequently manifests as backpressure and queue propagation through the service dependency graph. When a service becomes overloaded, requests may queue, reducing throughput at and around the overloaded service. These delays can then propagate upstream as services wait for responses from impacted dependencies \cite{wang2024AutothrottlePracticalBiLevel}. This often produces hotspots, where a single slow service causes queues to rapidly form and move outward \cite{gan2019anopensourcebenchmarksuite}.
    
    Propagated queues can affect network infrastructure in addition to microservices and their dependencies, which can both hasten their spread and worsen the subsequent delays \cite{dean2013TailScale}. Because queued requests continue to consume resources while waiting for execution (or timeout), an affected system can remain congested long after the initial slowdown has been fixed. Consequently, propagated queues can dramatically increase latency distributions across large portions of a system, and are difficult to mitigate \cite{dean2013TailScale, barroso2019DatacenterComputerDesigning}.

\subsection{System Instability}

    The speed at which small increases in latency can spread to large portions of a system can cause abrupt changes to the health of the system as a whole. Once queues, retries, resource contention, or failures begin to spread and reinforce each other, the availability and stability of services can rapidly decline \cite{barroso2019DatacenterComputerDesigning}. These instabilities are often the result of positive feedback loops, where mitigation actions create additional timing stress or oscillations.

    For example, attempts to recover from a host failure can overload the network, and the resulting congestion can appear as additional failures (see Sections \ref{sssec:coordination_and_reconfiguration} and \ref{ssec:feedback_and_control_loops}) \cite{rodrigues2005RobustServicesDynamic}. Similar effects occur in QoS-aware scheduling systems, where a missed deadline may cause subsequent deadline misses \cite{stankovic1995ImplicationsClassicalScheduling, delimitrou2014QuasarResourceefficientQoSaware}.
    
    As noted in Section \ref{ssec:queue_propagation}, instabilities and queues often form their own feedback loops, as accumulated queues require time to clear. In that time, persistent congestion continues to impact performance, and may spread to other parts of the system \cite{wang2024AutothrottlePracticalBiLevel}. The nonlinear relationship between timing effect propagation and system stability is a major reason for ongoing research into anomaly prediction, detection, and mitigation techniques \cite{yu2021MicroRankEndtoEndLatency, liu2021MicroHECLHighEfficientRoot, gu2023TrinityRCLMultiGranularCodeLevel}.

\subsection{Causal Ambiguity}

\begin{figure}[h]
        \centering
        \includegraphics[width=0.4\linewidth]{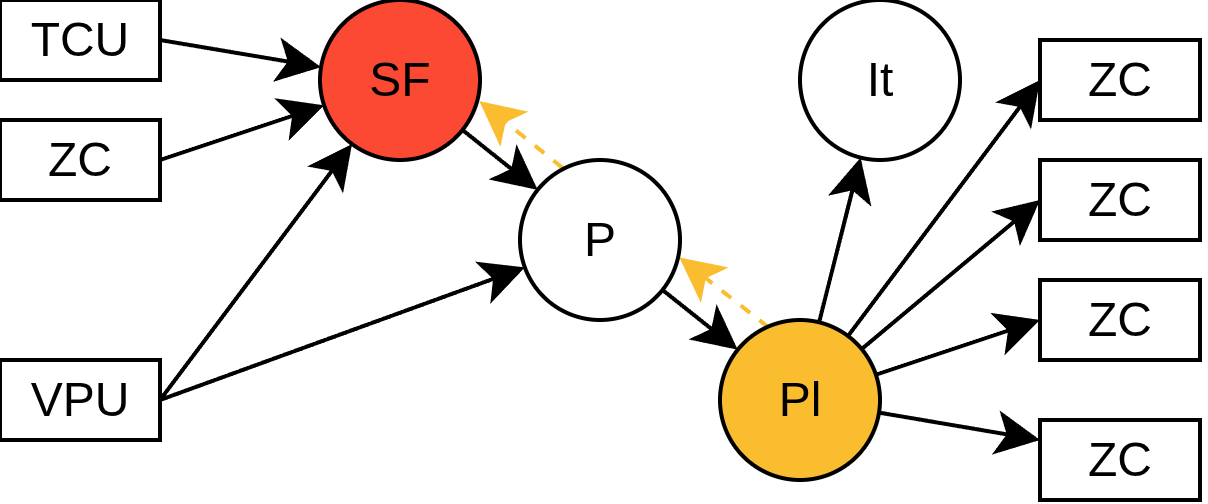}
        \Description{A reconstructed execution graph for a microservice-based SDV in which the source of an anomaly is several hops away from the service that triggers the anomaly detector.}
        \caption{The source of an anomaly (yellow) may be quite far from the anomaly detection trigger point (red). Tracing the source may be impossible without adequate instrumentation.}
        \label{fig:IMP_ambiguity}
    \end{figure}

    When timing effects propagate, the relationships between observed effects and their causes are often obscured. Anomalies spread through dependencies (Section \ref{ssec:dependency_graphs}) or middleware (Section \ref{ssec:feedback_and_control_loops}) may cause many services to appear anomalous simultaneously \cite{wang2020WorkflowAwareAutomaticFault}. The services exhibiting abnormal timing behavior are not necessarily responsible for the original disturbance. In Figure \ref{fig:IMP_ambiguity}, a slowdown in one service causes another to behave erratically, which triggers anomaly detection; without root cause analysis, the source of the anomaly is ambiguous.
    
    Limited observability in large or complex microservice-based systems compounds this ambiguity. Distributed tracing mechanisms may be unable to correctly reconstruct the causal relationships between dependent services, especially if those services are not adequately instrumented \cite{ashok2024TraceWeaverDistributedRequest}. Coarse-grained sampling windows can cause different anomalous patterns to appear indistinguishable, and tracing may itself be insufficient for identifying propagation paths if there is no direct invocation relationship between affected services (Sections \ref{ssec:resource_sharing} and \ref{ssec:amplification_effects}) \cite{cornacchia2026ObservabilityEatingYour, gu2023TrinityRCLMultiGranularCodeLevel}. In general, the symptoms reflect the propagation structures of the system rather than the source of the original disturbance \cite{lin2018MicroscopePinpointPerformance}. This makes root cause identification incredibly difficult when multiple modes of timing propagation exist.

\subsection{Nonlinear Timing Behavior}

    Interactions in microservice applications produce emergent behavior that is difficult to predict from the behavior of individual services. A single request may traverse many components, causing local delays to accumulate across execution paths and increasing end-to-end latency \cite{zhang2025NetworkAwareReliabilityModeling, almaruf2022UsingMicroserviceTelemetry}. As a system grows in complexity, so does the likelihood of amplification effects, errors, and misdiagnosis \cite{huang2025ConcurrencyAwareSelfDurationHierarchical}.

    The end-to-end latency of an application is rarely strictly linear, and local timing variabilities can produce disproportionately large effects on the system as a whole \cite{gan2019anopensourcebenchmarksuite}. This is particularly problematic in time-constrained systems, which are highly sensitive to small, local timing perturbations \cite{lee2008CyberPhysicalSystems}. Many of the mechanisms used to manage synchronization between concurrent executions (e.g., locks, interrupts, etc.) are brittle, and may increase the effects of downtime when failures propagate or cascade \cite{lee2008CyberPhysicalSystems, wang2023ComplexBehavioralInteraction}. As a result, end-to-end latency is a strong global indicator of application performance, but it should not be used as a predictive measure of its constituent parts.

\section{Timing Models}\label{sec:timing_models}

    This section surveys the major classes of formal models used to quantify and analyze timing behavior in distributed and microservice-based systems. Rather than present an exhaustive taxonomy of specific models, we instead focus on the contributions, assumptions, and admissible guarantees of families of models. The classes, as presented here, primarily differ in how they represent workload and service behavior, execution structures, concurrency, resource availability, uncertainty, and observable information. With this approach, we intend to give a general primer of existing timing models and to highlight their limitations when analyzing modern microservice-based systems.

\subsection{Analytical Models}

    \subsubsection{Network Calculus}

\begin{figure}[h]
            \centering
            \includegraphics[width=0.5\linewidth]{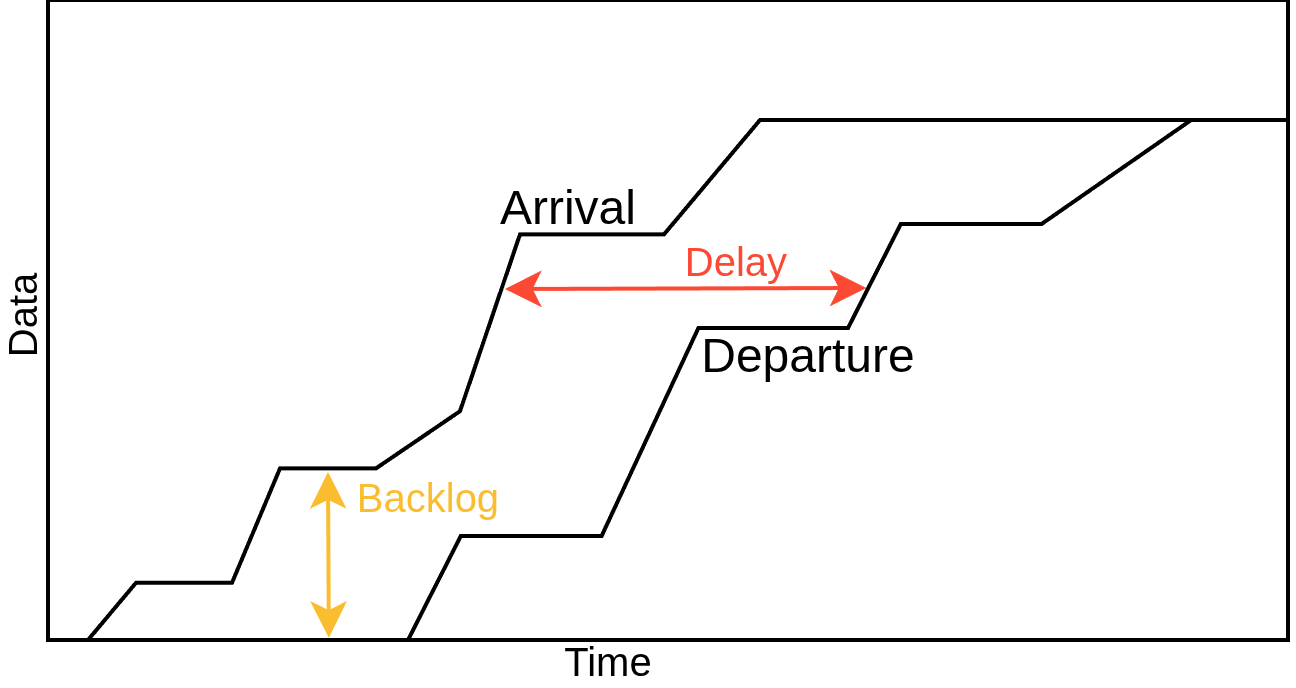}
            \Description{Arrival and Departure curves for a network. The delay (horizontal distance between the two curves) and backlog (vertical distance between the two curves) are highlighted.}
            \caption{Network calculus uses arrival and departure curves to model service behavior. Service backlog (yellow) and delay (red) can be calculated from the service curves.}
            \label{fig:MDL_network_calculus}
        \end{figure}

        Network calculus extends timing analysis beyond single-node systems by modeling how delay propagates across multi-stage service chains. It represents distributed execution as a composition of processing and communication stages, allowing end-to-end latency to be calculated from the bounded behavior of each constituent stage \cite{kannan2019GrandSLAmGuaranteeingSLAs}. In this framework, cumulative delay is analyzed through arrival curves, service curves, and resource constraints along the chain (Figure \ref{fig:MDL_network_calculus}), from which latency guarantees and scheduling policies can be derived \cite{leboudec1998ApplicationNetworkCalculus, leboudec2001NetworkCalculus}. This makes network calculus useful as a bridge between local component behavior and global timing behavior across distributed paths.
        
        However, network calculus depends on assumptions that may not hold in many modern software systems. This model generally relies on predictable request arrivals, stable service behavior, and sufficient resource availability. In microservice architectures, such assumptions are often weakened by dynamic workloads, asynchronous interactions, network failures, contention for shared resources, and virtualization overhead \cite{fan2016FormalAspectOrientedMethod, wen2024VirtualizationMicroserviceArchitecture, park2023AnalysisIPCANCommunication, pan2024SoftwareDefinedVehiclesModelBased}. As a result, network-calculus-based analysis can provide meaningful upper bounds, but guarantees may become conservative or inaccurate when applied to highly dynamic real-world systems \cite{kannan2019GrandSLAmGuaranteeingSLAs, zhou2018DeltaDebuggingMicroservice, fan2016FormalAspectOrientedMethod}.

    \subsubsection{Real-Time Calculus}
    
        Real-time calculus similarly allows timing analysis in distributed systems by modeling the propagation of timing constraints across chains of computation and communication. Like network calculus, it treats the system as a composition of interacting stages and derives end-to-end behavior from component-level abstractions based on arrival curves, service curves, and resource constraints \cite{kannan2019GrandSLAmGuaranteeingSLAs, thiele2000RealtimeCalculusScheduling}. This provides a structured method for reasoning about how local demand and availability interact and combine across a system.

        The guarantees provided by real-time calculus are strongest when request patterns are predictable and service availability is stable, though distributed systems frequently experience dynamic workloads, asynchronous communication, or communication failures \cite{fan2016FormalAspectOrientedMethod, wen2024VirtualizationMicroserviceArchitecture, park2023AnalysisIPCANCommunication, pan2024SoftwareDefinedVehiclesModelBased}. Thus, these guarantees may be overly conservative, or may not sufficiently represent the underlying system in dynamic or uncertain environments \cite{kannan2019GrandSLAmGuaranteeingSLAs, zhou2018DeltaDebuggingMicroservice, fan2016FormalAspectOrientedMethod}.

\subsection{Probabilistic Models}

    \subsubsection{Stochastic Traffic Patterns}

        Probabilistic timing models are often used when deterministic timing assumptions are either infeasible or too costly to guarantee. Rather than assuming fixed execution times or traffic patterns, these models represent request intervals, service times, and failures as stochastic processes \cite{stankovic1995ImplicationsClassicalScheduling}. Classically, these models assume that requests arrive according to a Poisson process, and that service times follow exponential or similar distributions. This enables quantities such as waiting time and throughput to be analyzed in a tractable manner \cite{thomas1976QueueingSystemsVolume, chou1977QueueingSystemsVolume}. More recent models extend this approach to include dynamic runtime conditions and probabilistic state transitions (see Section \ref{ssec:state-based_models}).

        Real-world microservice workloads frequently violate the assumptions of stochastic models, such as by following diurnal, bursty, noisy, or highly variable request patterns that are not well represented by simplified arrival distributions \cite{luo2022InDepthStudyMicroservice, wang2024AutothrottlePracticalBiLevel}. Similarly, long-tailed workload behavior can produce uneven resource demand in services \cite{harchol-balter2003SizebasedSchedulingImprove}. To better accommodate these behaviors, recent work has explored different methods for fitting statistical distributions to microservice traffic patterns \cite{luo2022InDepthStudyMicroservice}.

    \subsubsection{Tail Latency}

        In contrast to deterministic models, probabilistic models characterize timing behavior in terms of observed performance distributions rather than worst-case execution bounds \cite{ferraro2023TimesensitiveAutonomousArchitectures}. This makes them particularly useful for analyzing latency and jitter in systems where network fluctuations, queuing effects, or other instabilities introduce unavoidable uncertainty.
        
        In such settings, tail latency is especially important, because rare but extreme delays often dominate the perceived performance of complex distributed systems \cite{kannan2019GrandSLAmGuaranteeingSLAs, camilli2018FormalFrameworkSpecifying, fraccaroli2023TimingPredictabilityIPbased}. By focusing on high-percentile behavior rather than absolute worst-case bounds, probabilistic models provide a more realistic account of timing variability in large-scale systems where strict enforcement is difficult.

    \subsubsection{Probabilistic Guarantees}

        Probabilistic models typically derive their timing distributions from empirical observations, using runtime measurements to predict system behavior under realistic operating conditions \cite{yu2021MicroRankEndtoEndLatency}. This makes them well suited to environments where execution constraints cannot be strictly enforced, but it also means that the guarantees they provide are weaker than deterministic ones. Rather than ensuring that deadlines are always met, these models estimate the likelihood of delay under a given set of assumptions \cite{ferraro2023TimesensitiveAutonomousArchitectures}. Their accuracy may degrade when important confounding factors are omitted. This can include imperfect clock synchronization, bandwidth limitations, or node failures \cite{fan2016FormalAspectOrientedMethod, vanderaalst2012ProcessMiningOverview, choi2024EnhancingRecoveryPerformance, sperling2024ReducingCommunicationCost}.
        
        Many probabilistic approaches additionally depend on synchronized observations, centralized coordination, or periodic resource re-optimization, particularly when making online predictions; these models are often used alongside other techniques when strong guarantees are required \cite{fan2016FormalAspectOrientedMethod, do2024PerformanceAnalysisTraffic, teixeira2025DeterministicReliableSoftwareDefined}.

\subsection{State-Based Models}\label{ssec:state-based_models}
    
    State-based timing models represent distributed systems as a collection of states connected by transitions that describe how the system changes over time. They attempt to represent timing-related behaviors explicitly, and allow for timing constraints, concurrency, synchronization, and adaptations to be individually modeled and analyzed. This is a broad class of timing models, including automata-based models, Markov models, and multiple variants of Petri nets \cite{arcaini2015ModelingAnalyzingMAPEK, belusso2016StudyPetriNets, castro2002PracticalByzantineFault}.

\begin{figure}
        \centering
        \includegraphics[width=0.65\linewidth]{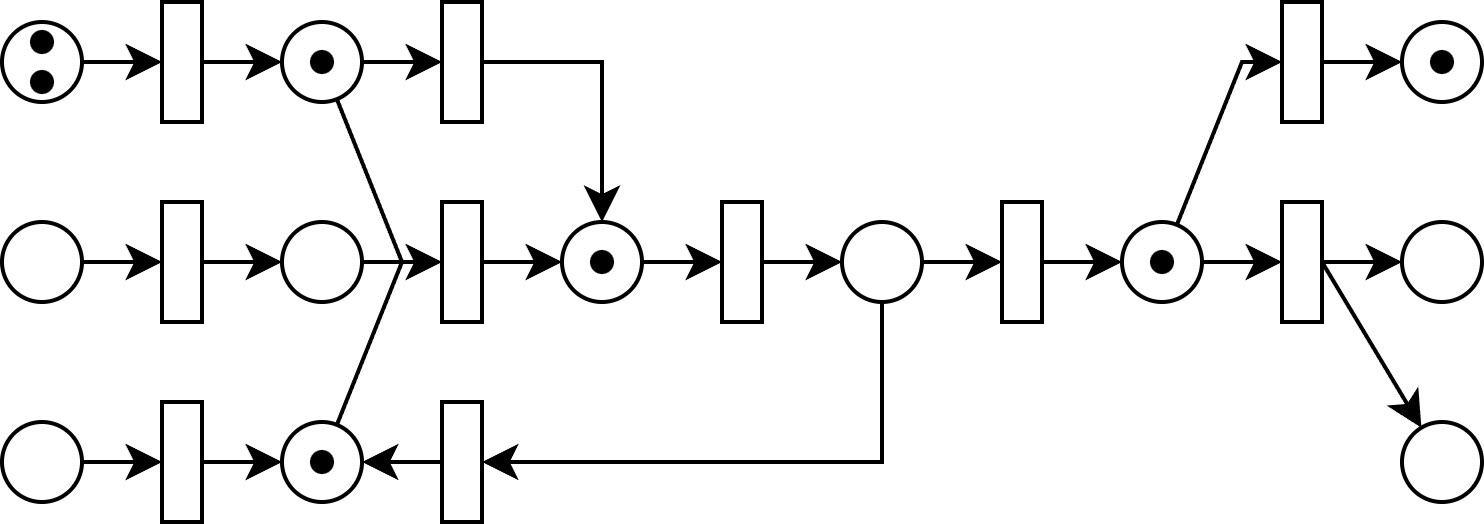}
        \Description{The example SDV deployment is rendered as a Petri net. Each state is drawn as a circle, and transitions are drawn as rectangles. Dots are placed at certain states to represent concurrency.}
        \caption{Petri nets can model concurrency in a system, but they become complex as the number of states and transitions increases.}
        \label{fig:MDL_petri_net}
    \end{figure}

    Petri net variants are commonly used to model microservice-based systems (e.g., Figure \ref{fig:MDL_petri_net} shows one such representation of the execution graphs presented in Figure \ref{fig:BCK_exec_graphs}), because they directly and clearly model concurrent and asynchronous events \cite{vanderaalst2012ProcessMiningOverview, belusso2016StudyPetriNets}. Many extensions exist which provide different ways of structuring behavior and constraints, with Workflow nets and Time Basic (TB) Petri nets among the most common for modeling the timing behavior of complex distributed systems \cite{camilli2018FormalFrameworkSpecifying, camilli2018HighlevelPetriNetbased, camilli2018ZonebasedFormalSpecification}.

    State-based models are particularly useful for analyzing schedulability in systems whose execution structure changes over time. For example, orchestration systems often transition between operating modes or resource configurations \cite{camilli2018DesignTimeRunTimeVerification, fan2016FormalAspectOrientedMethod}. These configurations can be represented as distinct states, allowing the model to determine whether all states are reachable.
    
    The main drawback of state-based approaches is scalability. Large systems with high levels of concurrency, dynamism, or probabilistic branching produce extremely large state spaces. These generally require abstractions or decomposition to remain tractable \cite{camilli2018FormalFrameworkSpecifying}. The challenge of balancing the expressiveness of the model against the complexity of its state space makes these models difficult to use in live environments, and they are frequently relegated to design-time system analysis \cite{camilli2018DesignTimeRunTimeVerification, dustdar2023DistributedComputingContinuum}.
    
\subsection{Deterministic Models}

    Many real-time systems require strong guarantees, and many deterministic models exist to analyze their schedulability and behavior. In the case of SDVs, real-time components are typically relegated to subsystems which themselves are tightly controlled. However, because those subsystems necessarily coexist alongside less timing-critical components, microservice-based SDVs must nonetheless guarantee their performance. For that reason, we must discuss deterministic models and their features.

    Real-time systems rely on explicit worst-case assumptions in order to provide bounds on execution timing; for example, to determine whether a set of tasks can meet their deadlines under a given scheduling policy \cite{stankovic1995ImplicationsClassicalScheduling, thiele2000RealtimeCalculusScheduling}. Execution bounds for safety-critical workflows are intentionally conservative, often assuming that multiple independent worst-case conditions may occur simultaneously \cite{stankovic1995ImplicationsClassicalScheduling}. When execution also includes communication delays, overhead, interference, or uncertainty, worst-case bounds are difficult to derive \cite{kannan2019GrandSLAmGuaranteeingSLAs, zhou2018DeltaDebuggingMicroservice, wen2024VirtualizationMicroserviceArchitecture}. This makes deterministic models hard to generalize to microservice environments, where unpredictability is often an unavoidable consequence of distributed architecture \cite{stankovic1995ImplicationsClassicalScheduling}.

    Worst-case execution timing (WCET) and schedulability analysis typically assume that task parameters, resource availability, and scheduling policies are known in advance. These assumptions fit relatively well in embedded or fully observable systems, but they are hard to guarantee in dynamic or heterogeneous environments. When any of these parameters are unknown, calculating task schedules may become computationally intractable \cite{camilli2018ZonebasedFormalSpecification, fan2013ModelingOptimizingResource, stankovic1995ImplicationsClassicalScheduling}.
    
    Some models, such as Schedule Abstraction Graphs (SAG), generalize to multiprocessor environments, allowing component-level models to compose \cite{thiele2000RealtimeCalculusScheduling, nasri2018ResponseTimeAnalysis}. However, full composability requires decomposition of global deadlines into component-level bounds across all execution stages, which may not be possible in dynamic systems \cite{kannan2019GrandSLAmGuaranteeingSLAs, fan2016FormalAspectOrientedMethod}. Even when each component is individually bounded, global behavior may be unpredictable due to amplification effects \cite{camilli2018ZonebasedFormalSpecification, stankovic1995ImplicationsClassicalScheduling, dean2013TailScale}. Thus, deterministic guarantees are most applicable in hard real-time SDV subsystems, but are likely incomplete for full microservice-based systems.

\section{Limitations of Current Timing Models}\label{sec:limitations}

    Timing guarantees are only admissible under a set of structural and epistemic assumptions about the system being analyzed. Many applications of the discussed models require bounded and consistently-observable system structure, resource allocation, time synchronization, or execution behavior, and often assume that these can be supervised from a sufficiently centralized perspective \cite{zhou2018DeltaDebuggingMicroservice, teixeira2025DeterministicReliableSoftwareDefined}. Classical timing models typically assume fixed or stable workloads, known scheduling policies, and bounded interference conditions \cite{leboudec2001NetworkCalculus, stankovic1995ImplicationsClassicalScheduling}. Analytical models similarly depend on the dependency structures, traffic patterns, and resource allocations being known in advance and remaining within operational bounds \cite{arcaini2015ModelingAnalyzingMAPEK, zhou2018DeltaDebuggingMicroservice, kannan2019GrandSLAmGuaranteeingSLAs}. Probabilistic models relax their deterministic assumptions, but still assume that the system behavior is sufficiently represented by statistical distributions \cite{camilli2018FormalFrameworkSpecifying, fan2013ModelingOptimizingResource}. In all cases, guarantees only remain meaningful so long as component-level behavior can be accurately modeled and extended to the larger system.

    These assumptions also imply strong visibility and control requirements. For example, in self-adaptive systems, formal reasoning is built around centralized MAPE-K feedback loops, where a controller monitors the system, maintains a record of system state, and plans corrective actions from a global vantage point \cite{arcaini2015ModelingAnalyzingMAPEK, pietrantuono2018RunTimeReliabilityEstimation}. In these models, admissibility assumes strict bounds on time synchronization, topology, resource contention, and availability of knowledge within a specified scope. When such guarantees are composed together, their validity depends on whether their relevant assumptions are also known, stable, and composable \cite{teixeira2025DeterministicReliableSoftwareDefined, kannan2019GrandSLAmGuaranteeingSLAs}.

    When the behavior of a software system is distributed, adaptive, or otherwise uncertain, the assumptions of these models begin to break down. Execution conditions in SDVs (and, more generally, in microservice-based systems) usually evolve during operation, making static design-time scheduling impractical \cite{fan2013ModelingOptimizingResource, fan2016FormalAspectOrientedMethod}. As such, these systems increasingly rely on adaptive controls to manage runtime decisions, which limits the applicability of models that assume stable or centralized execution models. The mechanisms intended to reduce global timing variability also tend to conflict with the strict timing requirements of individual components, further complicating the interpretation and composition of local guarantees \cite{sperling2024ReducingCommunicationCost}. Timing models based on static assumptions cannot fully capture dynamic system behavior, even when governed by a centralized controller \cite{fan2016FormalAspectOrientedMethod, whaiduzzaman2021ResilientFogIoTFramework}.

    As the scale and complexity of a distributed system increases, it becomes more difficult to maintain a coherent global view of system state. In practice, eliminating various forms of latency variability becomes less realistic than designing a tolerant system \cite{dean2013TailScale}. Because centralized coordination models cannot fully control the behavior of a distributed application (whose components are not fully observable), any guarantees derived under uncertainty are necessarily approximate \cite{kannan2019GrandSLAmGuaranteeingSLAs}. This prevents existing deterministic models from being broadly applicable to uncertain and partially-observable use-cases, such as SDVs.

    In general, each constituent part of a microservice-based system may be analyzable on its own by existing models, but those individual analyses cannot be composed into a system-scale model without making broad assumptions about its overall operating conditions or environment. The applicability of a specific timing model to a time-constrained microservice-based system depends on whether its structural, operational, or observability assumptions remain valid for the relevant execution path and operating conditions of the system. Because microservice-based systems are not typically static, fully observable, or deterministic, they exhibit nonlinear timing behavior that current models do not fully capture \cite{camilli2018ZonebasedFormalSpecification, camilli2021FormalSpecificationVerification, teixeira2025DeterministicReliableSoftwareDefined}.
    
\section{Conclusion}

    In this survey, we have examined timing variability in microservice-based systems using software-defined vehicles as a motivating example. Rather than framing timing behavior as an additive product of local execution, we have shown that it instead emerges from interactions between service dependencies, shared resources, and feedback loops. When local timing disturbances cross service, resource, or coordination boundaries, they can propagate and amplify into complex behaviors that affect the predictability and stability of the system as a whole. While existing timing models are useful for analyzing the constituent components of a system, these component analyses may not generalize or compose when the system is dynamic, uncertain, or partially observable. On their own, the system-level guarantees provided by existing models may be insufficient for complex microservice environments such as SDVs.

    As a result, timing guarantees in microservice-based SDVs should be derived with regard to clear assumptions about execution structure, shared resources, propagation paths, workload, runtime dynamism, and observability. By making these assumptions explicit, timing analyses and guarantees may become more reliable, particularly in time-constrained and highly dynamic distributed environments.

\begin{acks}

    Funded by the European Union. Views and opinions expressed are however those of the author(s) only and do not necessarily reflect those of the European Union or Chips Joint Undertaking. Neither the European Union nor Chips Joint Undertaking can be held responsible for them. The project is supported by the Chips Joint Undertaking and its members, including the top-up funding by the national Authorities of Austria, Germany, Italy, Denmark, France, Turkey, Finland, Spain, Portugal, Netherlands, Poland, Greece, Sweden, Romania, and the Czech Republic, under Grant Agreement Number 101194245 (Shift2SDV).
    
\end{acks}

\bibliographystyle{ACM-Reference-Format}
\bibliography{references}

\end{document}